\documentclass[preprint2]{aastex6}

\usepackage{amsmath,amssymb}
\def\arcsec{\hbox{$^{\prime\prime}$}}

\graphicspath{{./figs/}}

\shorttitle{Directly Observing the Galaxies Likely Responsible for Reionization}
\shortauthors{Livermore, Finkelstein \& Lotz}

\begin{document}

\title{Directly Observing the Galaxies Likely Responsible for Reionization}

\author{R. C. Livermore\altaffilmark{1} and S. L. Finkelstein}
\affil{The University of Texas at Austin, 2515 Speedway, Stop C1400, Austin, TX 78712, USA \vspace{-6mm}}
\and
\author{J. M. Lotz}
\affil{Space Telescope Science Institute, 3700 San Martin Drive, Baltimore, MD 21218, USA}

\altaffiltext{1}{rclivermore@utexas.edu}

\begin{abstract}
  We report a new analysis of the Hubble Frontier Fields clusters Abell 2744 and MACS 0416 using wavelet decomposition to remove the cluster light, enabling the detection of highly magnified ($>$50$\times$) galaxies a factor of 10$\times$ fainter in luminosity than previous studies. We find 167 galaxies at $z \gtrsim 6$, and with this sample we are able to characterize the UV luminosity function to $M_{\rm{UV}} = -12.5$ at $z \sim 6$, $-14$ at $z \sim 7$ and $-15$ at $z \sim 8$. We find a steep faint-end slope ($\alpha < -2$), and with our improved statistics at the faint end we reduce the fractional uncertainty on $\alpha$ to $<2\%$ at $z \sim 6 - 7$ and $4\%$ at $z \sim 8$. We also investigate the systematic uncertainty due to the lens modelling by using every available lens model individually and comparing the results; this systematic fractional uncertainty on $\alpha$ is $< 4\%$ at all redshifts. We now directly observe galaxies in the luminosity regime where some simulations predict a change in the faint-end slope of the luminosity function \citep{2013ApJ...766...94J,2015ApJ...807L..12O,2015MNRAS.453.1503B,2015arXiv151200563L} yet our results provide statistically very strong evidence against any turnover in the luminosity range probed, more consistent with simulations in which stars form in lower-mass halos \citep{2011MNRAS.410.1703F,2016arXiv160401314Y,2016arXiv160307729G}. Thus we find strong support for the extension of the steep luminosity function to $M_{\rm{UV}} = -13$ at $z>6$, consistent with the number of faint galaxies needed to reionize the Universe under standard assumptions.
\end{abstract}


\section{Introduction}

Since the installation of the Wide Field Camera 3 (WFC3) on the \emph{Hubble Space Telescope} (\emph{HST}), blank-field studies have discovered increasing numbers of faint, high-redshift galaxies. The faint end of the rest-frame ultraviolet (UV) luminosity function is steep \citep[$\alpha \sim -2$;][]{2011ApJ...737...90B,2015ApJ...803...34B,2013MNRAS.432.2696M,2015ApJ...810...71F}, implying that the UV luminosity density -- and therefore the ionizing photons responsible for the reionization of the Universe -- are dominated by faint galaxies. However, in order for galaxies to produce sufficient ionizing radiation to power reionization, one must integrate the luminosity function to at least a factor of 100 in luminosity below the direct observational limits of \emph{HST}, to an absolute UV magnitude of $M_{\rm{UV}} = -13$ \citep[assuming an escape fraction $f_\mathrm{esc} = 13 - 20$\%, Lyman continuum photon production efficiency of
$\log_{10} \xi_\mathrm{ion} = 53.14\,(\mathrm{Lyc\, photons\,s}^{-1}\,\mathrm{M}_{\odot}^{-1}\,\mathrm{yr})$ and clumping factor $C = 3$]{2015ApJ...810...71F,2015ApJ...802L..19R}. It is unclear that the luminosity function should continue unbroken to this magnitude; simulations indicate that it flattens at some magnitude $M_\mathrm{UV} > -17$. This flattening is caused both by the suppression of star formation in low-mass galaxies by heating from the ionizing background and by inefficient cooling of gas at low metallicities, such that at very low mass ($< 2 \times 10^8$M$_{\odot}$) not all halos contain stars \citep[e.g.][]{2011MNRAS.410.1703F,2013ApJ...766...94J,2015ApJ...807L..12O,2015arXiv151200563L,2016arXiv160401314Y,2016arXiv160307729G}. A flattened luminosity function at high redshift is also implied by observations of local dwarf galaxies, which should be $\sim 100\times$ more abundant if the luminosity function continues to rise beyond $M_\mathrm{UV} \sim -14$ at $z = 7$ \citep{2015MNRAS.453.1503B}.

If the luminosity function does flatten, this presents a challenge for models of reionization. However, the turnover, if it exists, lies beyond the nominal reach of \textit{HST}. The Hubble Frontier Fields (HFF) program (PI: Lotz) aims to approach this problem by observing massive galaxy clusters that efficiently gravitationally lens background galaxies that serendipitously lie along the line of sight, enabling the detection of faint galaxies whose intrinsic magnitudes lie beyond the \emph{HST} limits. However, work on the HFF clusters has been hampered by the difficulty of detecting faint galaxies in crowded cluster fields, and thus has been limited to characterising the UV luminosity function to $M_\mathrm{UV} \le -15$ at $z \sim 6$ \citep{2014ApJ...786...60A,2015ApJ...814...69AA,2015ApJ...800...18A,2015ApJ...799...12I} or $M_\mathrm{UV} \le -17$ at $z \sim 8$ \citep{2015ApJ...814...69AA, 2015ApJ...799...12I,2015AA...575A..92L}.  These previous studies have not probed the regime where theory implies the luminosity function may turn over.

In order to overcome these difficulties and fully exploit the lensing magnification, it is necessary to model and subtract the foreground galaxies and intracluster light. The {\sc ASTRODEEP} collaboration has recently had promising results using {\sc Galfit} to model the foreground light \citep{2016arXiv160302460M,2016arXiv160302461C}. As another approach to this problem, we have developed a new technique using wavelet decomposition to separate the images onto different physical scales and isolate the small, faint objects. With this technique, we are able to detect galaxies closer to the critical line where intracluster light and the high density of foreground galaxies usually hamper the detection of faint sources. In this paper, we apply this method to the first two Hubble Frontier Fields, Abell 2744 and MACSJ0416.1-2403 (hereafter MACS 0416), which have updated lens models incorporating the data obtained as part of the Frontier Fields program.

This paper is organized as follows: in \S\ref{sec:data} we introduce the data, our wavelet decomposition method is detailed in \S\ref{sec:wv}, the sample selection is discussed in \S\ref{sec:catalog}, and we compute the luminosity function in \S\ref{sec:lf}. We discuss the results in \S\ref{sec:disc} and present our conclusions in \S\ref{sec:conc}. Throughout, all magnitudes are given in the AB system and we adopt the same $\Lambda$CDM cosmology used in the lens modelling, with $H_0 = 70 $km\,s$^{-1}$\,Mpc, $\Omega_{\Lambda} = 0.7$ and $\Omega_M = 0.3$.

\section{Data}
\label{sec:data}

The data comprise two lensing clusters observed with the Advanced Camera for Surveys (ACS) and WFC3 on \emph{HST} as part of the HFF program. The clusters are observed in seven filters as described in Table \ref{tab:data}, and the drizzled science images and inverse variance weight maps were publicly released on the HFF website\footnote{\url{http://archive.stsci.edu/pub/hlsp/frontier/}}. We use the v1.0 releases with a pixel scale of 0.06\arcsec/pixel. For the ACS images, we use the images calibrated with a new `self-calibration' approach, which uses the science images to improve the dark subtraction (Anderson et al. in prep), and for the WFC3 images we use those corrected for time-variable background.

In order to measure accurate photometry from images at different wavelengths within the same aperture, we match all of the images to the same point-spread function (PSF). The process used for PSF-matching is the same as that used by \citet{2010ApJ...719.1250F}; briefly, all images are matched to the largest PSF, which is the $H_{160}$ band with full width at half maximum (FWHM) $\sim 0.18''$. We first measure the PSF in each band by stacking all of the stars in the field that do not have bright neighbouring sources. Kernels for each band are generated using the IDL {\sc deconv\_tool} routine, which performs an iterative process based on the PSF in each band and in the reference $H_{160}$ band. The images are then convolved with these kernels to produce PSF-matched images. We increase the number of iterations until the curve-of-growth of stars in the PSF-matched images match those in the $H_{160}$ band to within $\pm 1\%$ in circular apertures with a radius of 0.4''.

\begin{figure}
  \centering
  \includegraphics[width=0.5\textwidth]{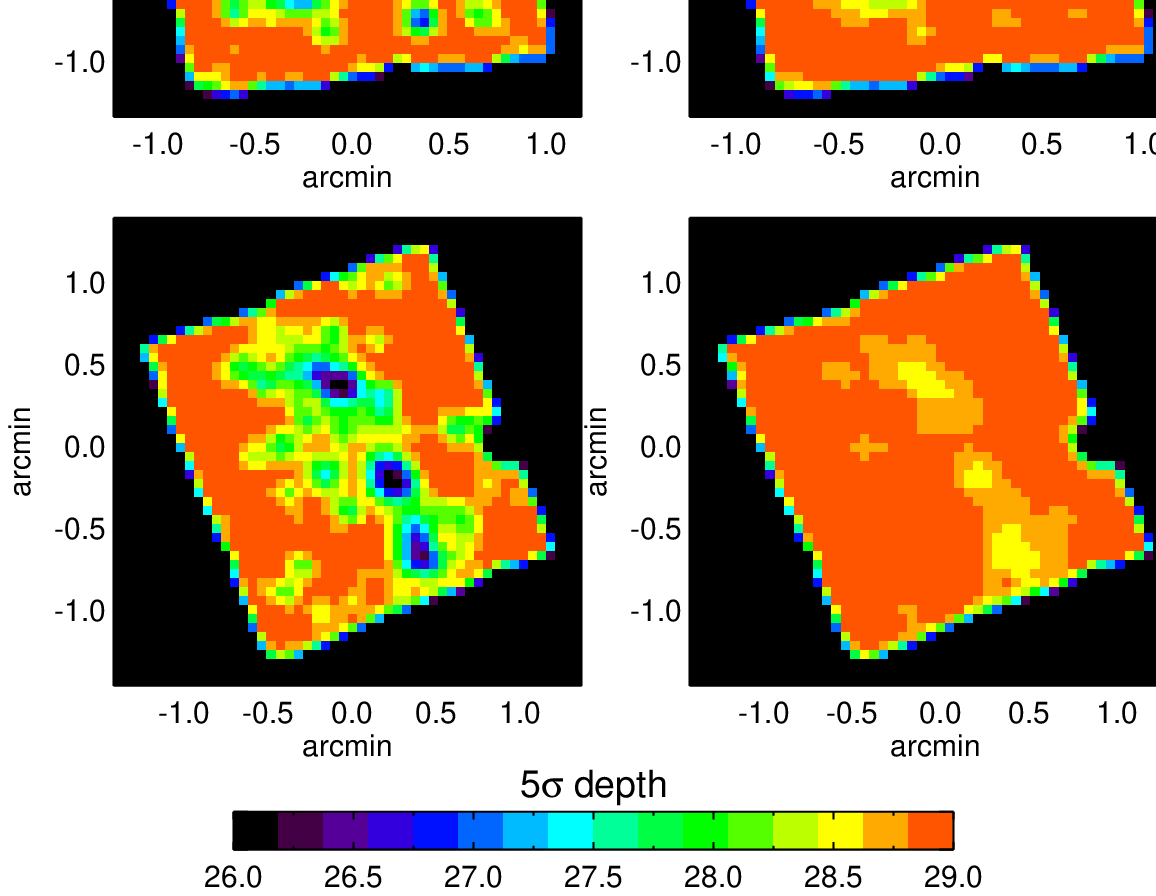}
  \label{fig:depthdrz}
  \caption{5$\sigma$ depth maps of Abell 2744 (\emph{top}) and MACS 0416 (\emph{bottom}), measured with $10^6$ randomly-placed 0.4'' apertures. Prior to wavelet decomposition (left), the depth is hampered in the central regions of the clusters due to bright foreground sources and intracluster light. After wavelet decomposition (right panels), the depth is significantly improved, especially in the central regions of the clusters.}
\end{figure}

To quantify the depth of the images, we measure fluxes in 0.4'' apertures in $10^6$ random positions. The 1$\sigma$ depth is measured from the negative side of the flux distribution, in order to exclude the contribution of sources in the image. We list the mean 5$\sigma$ depths in each filter for the two cluster fields in Table \ref{tab:data}. However, the depth varies substantially across these fields due to the bright foreground cluster galaxies and associated intracluster light (ICL). We therefore also measure the depth as a function of position by dividing the fields into 52$\times$52 regions and separately measuring the depth as above for each position. The resulting depth maps are shown in Figure \ref{fig:depthdrz}; the depth is substantially reduced in the center of the cluster where foreground sources and ICL dominate, making us much less sensitive to background high-redshift galaxies in these regions.

\begin{table}
\begin{deluxetable}{l c c c c}
  \tablecaption{Summary of the \emph{HST} data on the first two Hubble Frontier Fields clusters, Abell 2744 and MACS 0416 \label{tab:data}}
  \tablecolumns{5}
  \tablewidth{0pt}
  \tablehead{
    \colhead{Filter} & \multicolumn{2}{c}{Abell 2744} & \multicolumn{2}{c}{MACS 0416} \\
    & \colhead{Orbits} & \colhead{5$\sigma$ Depth} & \colhead{Orbits} & \colhead{5$\sigma$ Depth}}
  \startdata
  F435W & 24 & 29.0 & 21 & 29.2 \\
  F606W & 14 & 29.0 & 13 & 29.3 \\
  F814W & 46 & 29.3 & 50 & 29.5 \\
  F105W & 24.5 & 29.1 & 26 & 29.2 \\
  F125W & 12 & 29.0 & 13.5 & 28.9 \\
  F140W & 10 & 29.0 & 11 & 28.9 \\
  F160W & 24.5 & 28.8 & 26.5 & 28.8 \\
  \enddata
  \tablecomments{The depth is measured in randomly positioned 0.4\arcsec circular apertures in the original reduced images. We list the mean across the image.}
\end{deluxetable}
\end{table}

\section{Wavelet decomposition}
\label{sec:wv}

The depth of the images is limited by the presence of foreground cluster galaxies and ICL. To maximize our sensitivity to background high-redshift galaxies, we need to model and remove this light from the images. We can do this by isolating structures on different physical scales; we are interested in isolating $z > 6$ galaxies, which are much smaller on the sky than elliptical cluster galaxies at $z = 0.308$ and $0.396$ (for Abell 2744 and MACS 0416 respectively) or the cluster-scale ICL.

One method of decomposing images into different physical scales is with the wavelet transform \citep{2002aida.book.....S}. This tool has been widely used in astronomy for purposes such as de-noising images \citep{2002PASP..114.1051S}, analysing the clustering of galaxies \citep{1993ApJ...409..517S,1995AA...298....1E,2005ApJ...634..744M} and detecting non-Gaussianity in the Cosmic Microwave Background \citep{2004AA...416....9S,2004ApJ...609...22V}.

Briefly, the wavelet transform allows us to decompose an image into its components on different physical scales. Thus, we can isolate structures on large scales - such as the cluster galaxies and ICL - and remove them, allowing objects on smaller scales to be identified more easily.

The detailed process for calculating the wavelet transform of an image $I$ on $n$ scales is as follows:

\begin{enumerate}

\item To initialize the procedure, we set the scale $j=0$ and the input image $I_{j,\rm{in}} = I$.

\item We define the filter $h_j$ with which to convolve the image. We use a 1D filter $h$ derived from the $B_3$-spline \citep{2002aida.book.....S,2006AA...446.1191S}, defined as

\begin{equation}
  h_{1D} = \frac{1}{16}\left[1,4,6,4,1\right].
\end{equation}

For a scale $j > 0$, the size of the filter $h^j_{1D}$ is grown to $k = 4\times 2^j + 1$ elements, with the array elements all set to zero except for the elements $0, k/2 - 2^j, k/2, k/2+2^j$ and $k-1$, which are populated by the values of $h_{1D}$. Thus, the values are spaced by gaps that increase with higher values of $j$. This method of convolution is therefore known as an \textit{\`{a} trous}, or ``with holes.''

\item Each row of $I_{j,\rm{in}}$ is convolved with the filter $h^j_{1D}$ to produce an image $I_{j,\rm{out}}$.

\item Each column of $I_{j,\rm{out}}$ is convolved with the filter $\left(h^j_{1D}\right)^T$.

\item The wavelet transform of $I$ on scale $j$, is $w_j = I_{j,\rm{in}} - I_{j,\rm{out}}$.

\item We increment $j$ by 1 and set $I_{j,\rm{in}} = I_{j-1,\rm{out}}$, then repeat from step 2 until $j=n$.

\end{enumerate}

\begin{figure*}
\centering
  \includegraphics[height=\textwidth, angle=90]{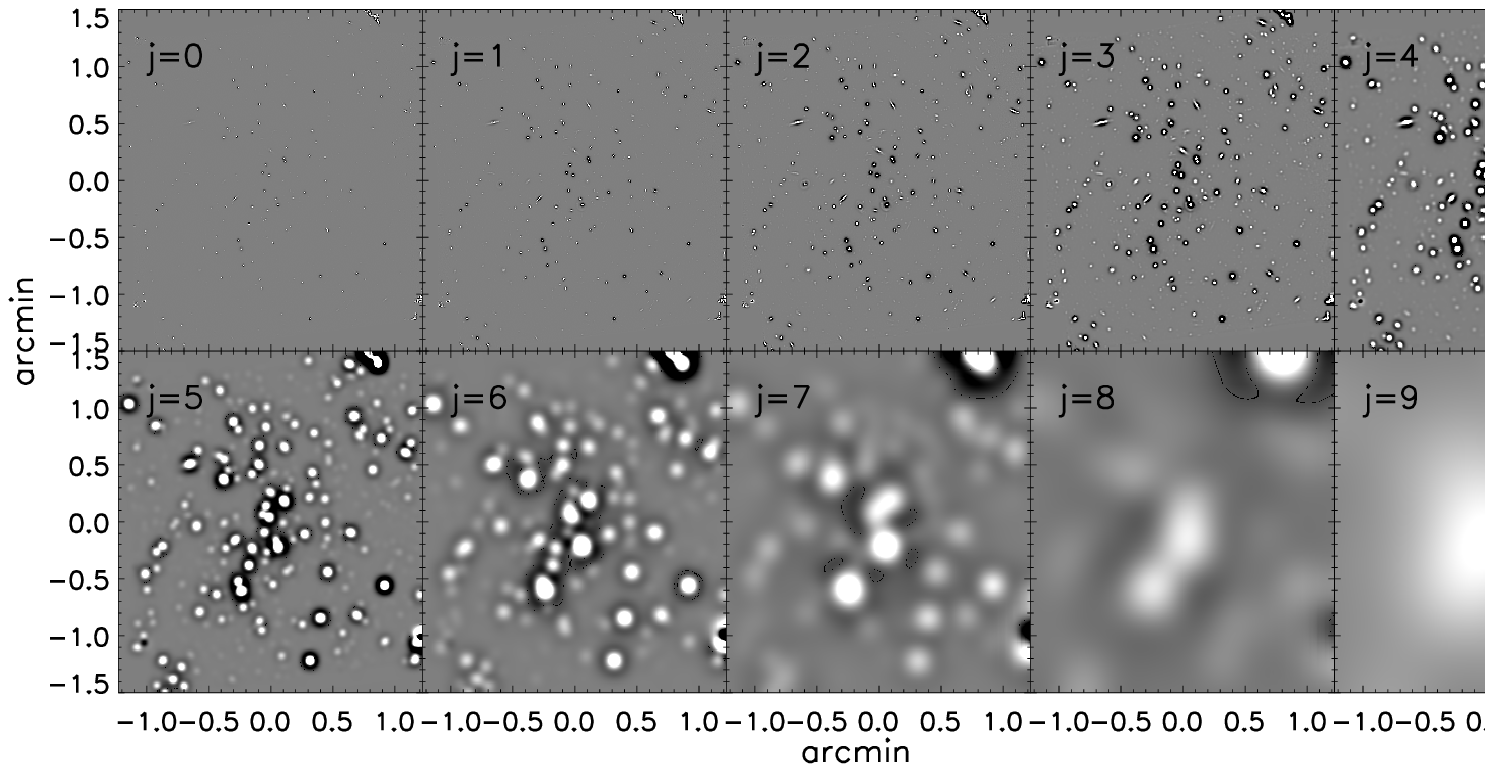}
  \caption{The $H_{160}$-band image of Abell 2744 decomposed into 10 scales using the method described in the text. On the largest ($j=9$) scale, we see the intracluster light. Small background galaxies appear on the smallest ($j < 5$) scales.}
  \label{fig:wvim}
  \end{figure*}

The result of this process is a series of $n$ images, representing the original image $I$ on increasing physical scales. We use $n=10$ and illustrate the decomposition of the Abell 2744 cluster in Figure \ref{fig:wvim}. We note that the original image $I$ can be exactly reproduced by summing all of these decomposed images.

In order to isolate small features, we need to use the decomposed images to model and then subtract the larger features in the field. To do this, we carry out the following procedure:

\begin{enumerate}

\item We run {\sc Source Extractor} ({\sc SExtractor}; \citealp{1996AAS..117..393B}) on the wavelet-transformed image $w_j$ to locate structures in the image.

\item Peaks in the wavelet-transformed images tend to be surrounded by negative rings. To find these, we multiply the image $w_j$ by $-1$ and run {\sc SExtractor} on the negative image.

\item Using the segmentation maps from steps (i) and (ii), we match the negative rings to their positive counterparts and create a merged segmentation map.

\item For the $j=9$ (largest) scale, each object is assigned an ID number of 9. For scales $j < 9$, we match each object to those identified on larger scales. Anything that is identified on a larger scale is assigned an ID number corresponding to the largest scale on which it was first detected. Any object not already detected is assigned the ID $j$. The resulting ID number of each pixel on each scale is stored in a master segmentation map $s_j$. 

\item Steps 1 - 4 are repeated for each scale $j=9,8,7...0$.

\item We now create a model of the cluster on each scale $j$, an image $m_j$ of the same dimensions as the original image $I$. Each pixel $x,y$ is defined as

  \begin{equation}
    m_j\left[x,y\right] = \sum\limits_{i=j}^{n-1} w_i\left[x,y\right] + \sum\limits_{i=0}^{j-1} \begin{cases} w_i\left[x,y\right] & \mbox{if } s_i\left[x,y\right] \ge j \\
      0 & \mbox{if } s_i\left[x,y\right] < j \end{cases}
  \end{equation}

\item The cluster models $m_j$ are subtracted from the original image $I$, resulting in a set of $n$ residual images. We note that the final model $m_n$ is exactly equal to $I$, and thus leaves residuals of zero.
  
\end{enumerate}

\begin{figure*}
\centering
\includegraphics[width=0.49\textwidth]{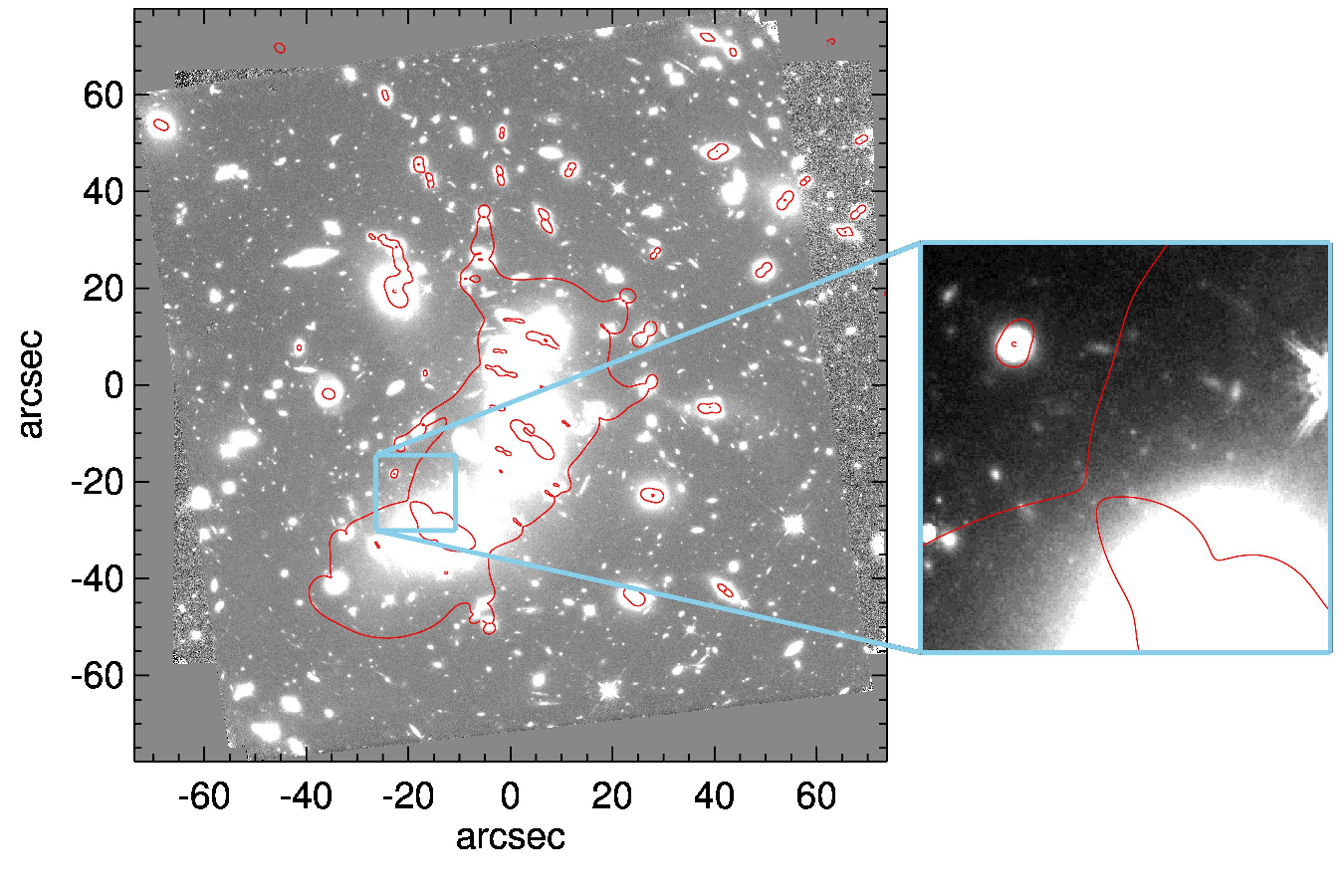}
\includegraphics[width=0.49\textwidth]{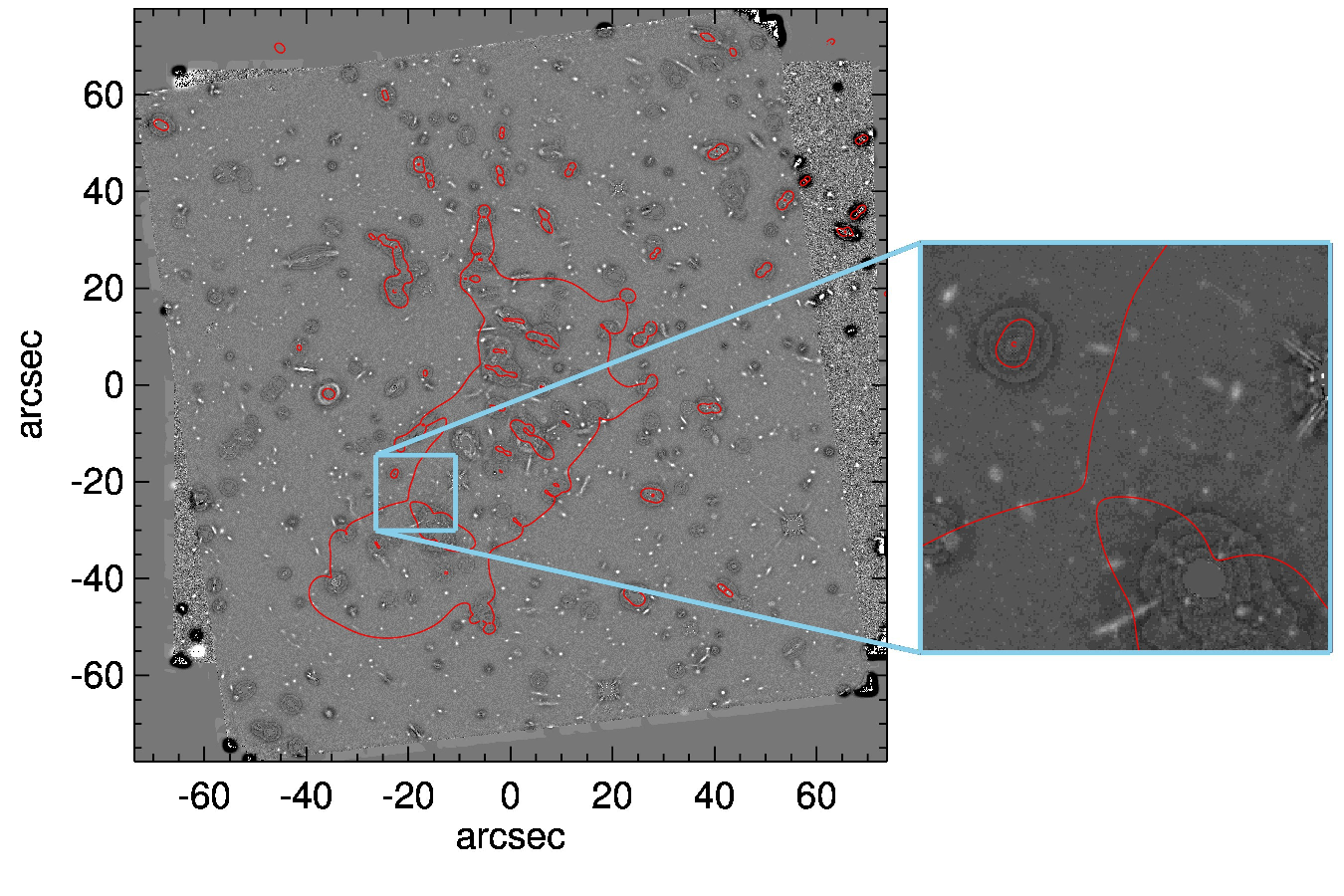}\\
\includegraphics[width=0.49\textwidth]{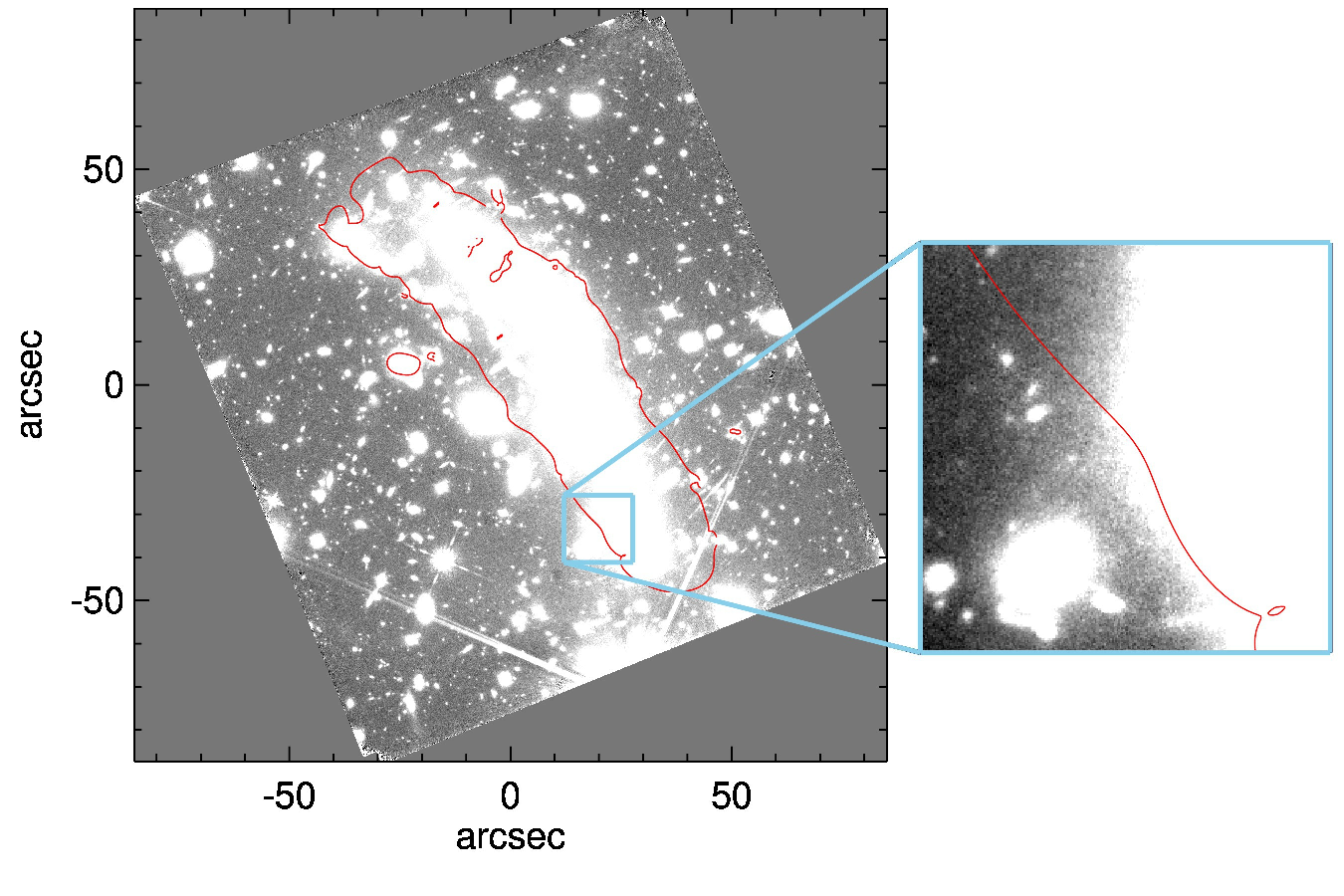}
\includegraphics[width=0.49\textwidth]{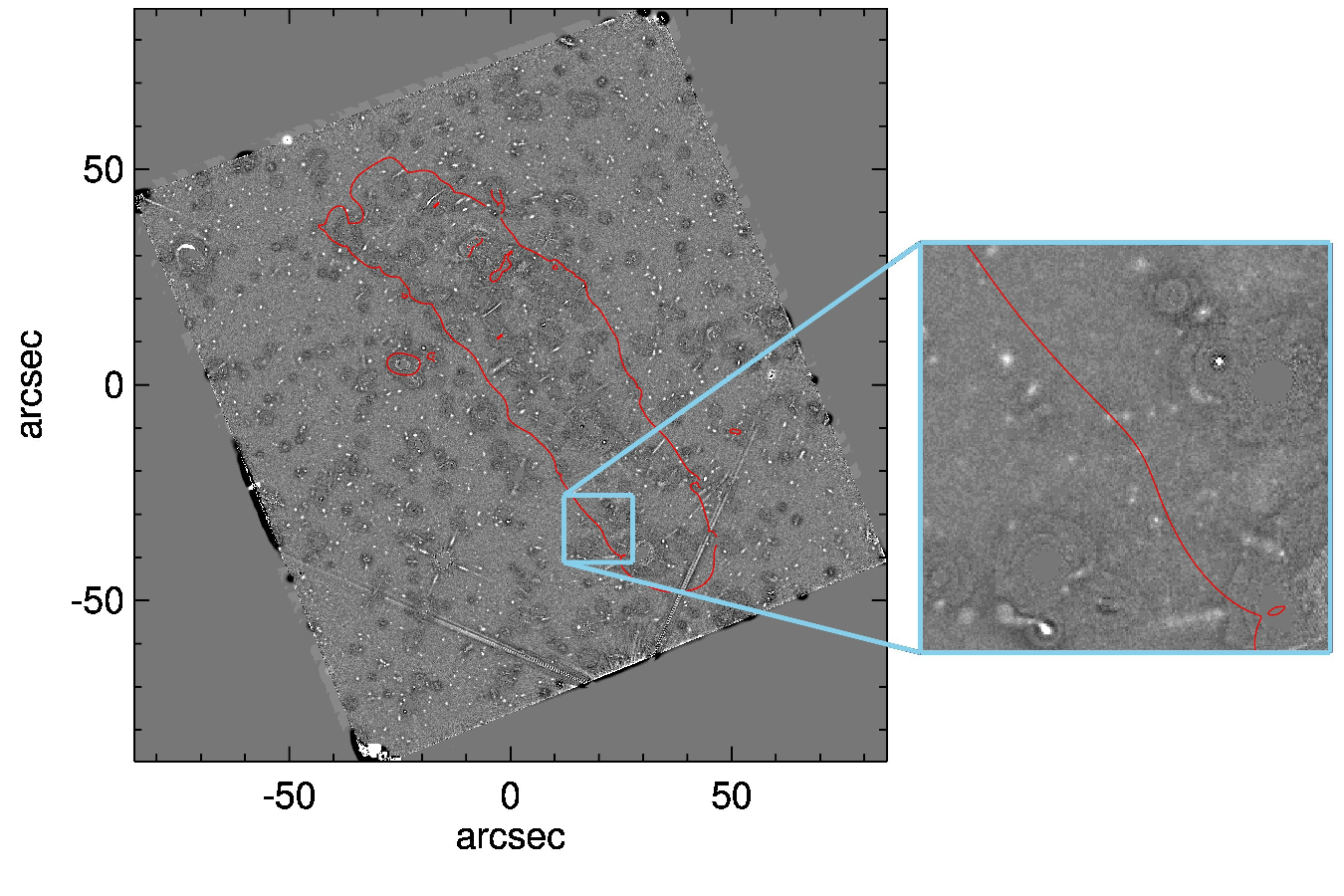}
\caption{$H_{160}$-band images of Abell 2744 (top) and MACS 0416 (bottom) before (left) and after (right) wavelet decomposition, shown with the same scaling. As a guide, the critical line at $z = 7$ is shown in red. The zoomed in regions (also shown with the same scaling) illustrate some of the faint sources that are obscured by cluster light in the original images but are easily visible after wavelet decomposition.}
\label{fig:wvims}
\end{figure*}

This procedure is carried out independently for each ACS and WFC3 band. The resulting $H_{160}$-band wavelet-subtracted images on scale $j=4$ are shown in Figure \ref{fig:wvims}.\\
\\

\subsection{Testing the wavelet decomposition method}
\label{sec:wvtest}

We carry out a number of tests to determine a) which of the residual images we should use for object detection, b) to what extent the wavelet decomposition procedure improves our detection efficiency, and c) whether we can measure reliable photometry from the residual images. As we need to determine the effect of the wavelet subtraction process on the detection and measurement of galaxies, we do this with a simulation in which we add fake galaxies to the data and attempt to recover them.

\begin{figure*}
  \centering
  \includegraphics[width=\textwidth]{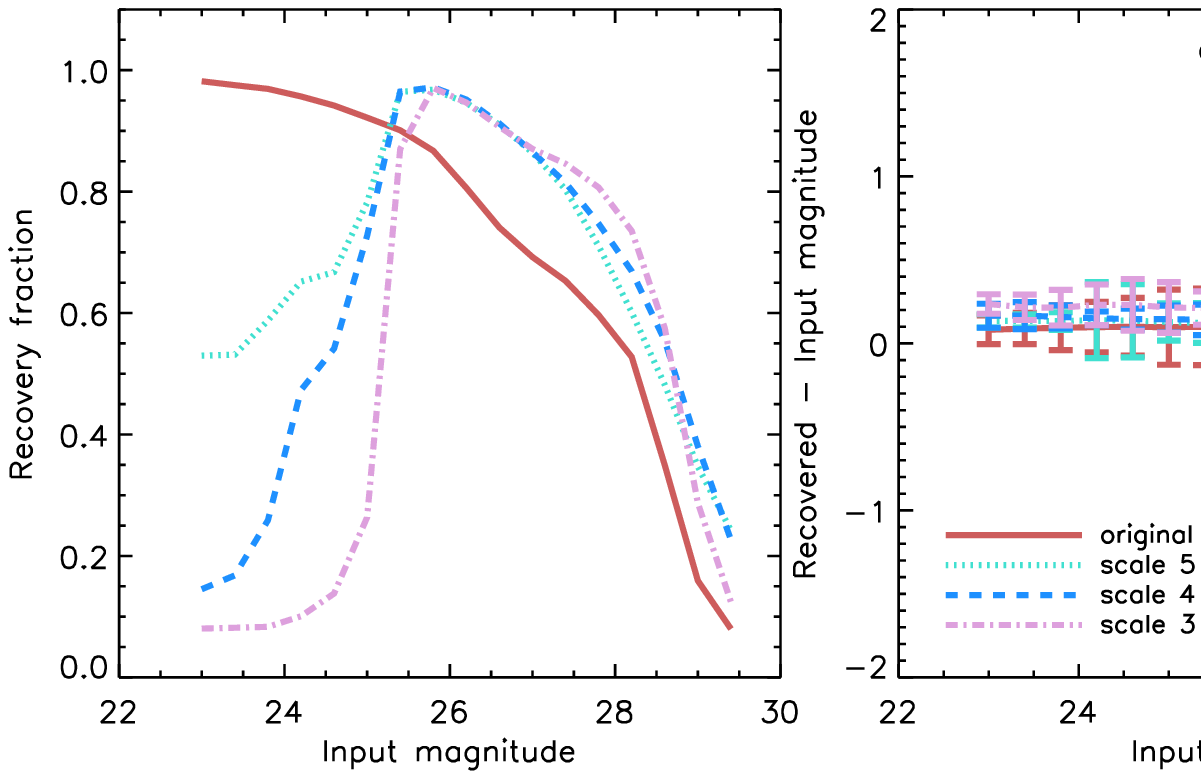}
  \caption{Results from simulations of fake source recovery in original and wavelet-subtracted images. We test all scales, but for clarity only scales $j = 3 - 5$ are shown. \emph{Left:} The total fraction of fake sources recovered as a function of input $H_{160}$-band magnitude. Recovery of faint sources is improved with the wavelet subtraction method. Bright sources are lost as they are subtracted in the wavelet decomposition process; to ensure we can detect bright as well as faint sources, we therefore use both the original and wavelet-subtracted images for source detection. \emph{Center:} Recovered - input magnitudes of fake sources as a function of input magnitudes for sources detected in the original and wavelet-subtracted images, but all photometry measured in the original image. The accuracy of photometry is noticeably affected when the objects are detected in images subtracted to scale 3. \emph{Right:} As center panel, but with photometry measured in the wavelet-subtracted images. Here, we recover systematically fainter magnitudes as part of the light of some of the galaxies is subtracted during the wavelet subtraction procedure.}
  \label{fig:wvsim}
\end{figure*}

We first generate fake point sources with a range of input magnitudes, convolve them with the $H_{160}$-band PSF and add them to the $H_{160}$-band image in random positions. We then compute the wavelet transform as detailed above, construct models on each scale $j=1...9$ and run {\sc SExtractor} on both the input image and the residual images on each scale. In Figure \ref{fig:wvsim}, we compare the results. As the left panel of Figure \ref{fig:wvsim} shows, the wavelet images are useful for detecting sources at $m > 26$; brighter than this, all or part of the galaxy is removed in the wavelet-subtraction process. Although high-redshift galaxies brighter than this limit are rare, we therefore need to use both the original and wavelet-subtracted images for detection to ensure that we find any that may exist. We also need to ensure that we can measure accurate photometry when using the wavelet-subtracted images for detection. The center panel of Figure \ref{fig:wvsim} shows that photometry measured in the original image for sources detected in the wavelet-subtracted images (scales 3-5) is of comparable accuracy to that of sources detected in the original image (but note that more sources are included for the wavelet-subtracted images at the faint end). If we both detect and measure sources in the wavelet-subtracted images, we find that in the range where sources can be efficiently recovered ($m > 26$) the images subtracted on scales 4 and 5 have measurement errors comparable to sources measured in the original image (again, over more sources overall). However, there is a systematic shift towards fainter magnitudes: this is because objects that are partially removed in the wavelet-subtraction process or lie close to residuals are still included.  Therefore, we only use the wavelet photometry when it is within 1$\sigma$ of the photometry on the original image (using the wavelet-subtracted image for detection in both cases). This results in accurate recovered photometry on scale $j=4$ while maximizing the signal-to-noise by using the wavelet-subtracted image where possible. If we use scale $j=3$, the recovered flux is reduced even when measuring in the original image because oversubtraction of galaxy light causes the Kron apertures to be drawn too small. Scale $j=4$ is therefore our preferred scale on which to model and subtract the cluster light, and this scale is used throughout the remainder of this paper.

We also note that detection in the wavelet-subtracted images is hampered for bright $\left(H_{160} < 25\right)$ galaxies as they get removed when we perform the subtraction to smaller scales. Although we do not expect to find any high-redshift galaxies in this regime, we nonetheless choose to combine catalogs from the wavelet-subtracted and original images in order to ensure that we include potential bright or extended sources.

\section{Source Selection}
\label{sec:catalog}

We detect galaxies in these fields using {\sc SExtractor} in dual image mode, where one  detection image is used to measure photometry in each of the seven available \emph{HST} filters. In order to maximise our sensitivity to faint high-redshift galaxies at a range of different colors, we use eleven different detection images: each of the four \emph{HST}/WFC3 filters $\left( Y_{105}, J_{125}, JH_{140} \mathrm{~and~} H_{160}\right)$, plus seven stacked combinations of these images: $H_{160} + JH_{140} + J_{125} + Y_{105}$, $H_{160} + JH_{140} + J_{125}$, $H_{160} + JH_{140}$, $JH_{140} + J_{125} + Y_{105}$, $JH_{140} + J_{125}$, $J_{125} + Y_{105}$, and $JH_{140} + J_{125}$.

We carry out this process with detection in both the original PSF-matched images and the wavelet-subtracted ones, for a total of 22 detection images. In all cases, photometry is measured from the original, PSF-matched, images in all seven filters as well as from the wavelet-subtracted images. We use the wavelet-subtracted photometry where it is within 1$\sigma$ of the photometry from the original image in every band, because the signal-to-noise is higher, especially in the central regions of the clusters. Where the photometry differs (i.e. where some part of the galaxy has been subtracted in the wavelet decomposition process), we use the photometry from the original image. We also visually inspect every galaxy to ensure that no part of the galaxy has been subtracted; we find that all partially-subtracted galaxies have photometry that differs by $>1\sigma$ between the original and wavelet-subtracted images, so the original image photometry is used.

We then combine our 22 catalogs into a single master catalog, selecting unique sources within a matching radius of 0.2''. We measure photometry in the small Kron aperture flux, with the {\sc SExtractor} parameter PHOT\_AUTOPARAMS set to 1.2 and 1.7, (although see Section \ref{sec:vis} for corrections to some sources). We remove some sources based on a mask created from the weight map, which excludes noisy regions at the edges of the field. We derive the aperture correction to the total flux for each source from the $H_{160}$ band, using the ratio between the flux obtained with a larger aperture (PHOT\_AUTOPARAMS = 2.5,3.5) and that described above. The same correction is then applied to the flux and flux errors in all filters. We then correct for galactic extinction using the E(B-V) value for each field from \citet{2011ApJ...737..103S} and a reddening curve appropriate to each filter \citep{1989ApJ...345..245C}.

\subsection{Photometric redshifts}
\label{sec:photoz}

To select high-redshift galaxies from our master catalog, we use a photometric redshift fitting technique. This has the advantage over color selection techniques that it uses all of the available photometry and results in more precise estimates of the error on the resulting redshifts. The redshift fitting code we use is {\sc EAZY} \citep{2008ApJ...686.1503B}, with an updated set of templates that include contributions from emission lines.

We select high-redshift candidates from the master catalog using the full redshift probability (P($z$)) distribution as well as the best-fit redshift and the detection significance in the WFC3 filters. In our high-redshift selection criteria, we follow previous work on the unlensed CANDELS fields by \citet{2015ApJ...810...71F} in order to ensure our results are comparable. Therefore, to be selected, a galaxy must meet all of the following criteria:

\begin{enumerate}
\item The best-fit redshift must be $> 4$.
\item The integral of the main peak of the P($z$) distribution must be $>70\%$.
\item The integral of the P($z$) distribution in one of the $z \ge 6$ redshift bins (where a bin is defined as an integer $\pm 0.5$) must be $> 25\%$.
\item The flux must be $> 3.5\sigma$ in at least two WFC3 bands, or $> 5\sigma$ in the $H_{160}$ band, where we define $1\sigma$ in a position-dependent fashion using the depth maps described in Section \ref{sec:data}.
\end{enumerate}

These criteria ensure that the candidates are firm detections with well-defined photometric redshifts, and that the high-redshift solution is strongly preferred. To allocate each candidate to a redshift bin $z_{\rm{bin}}$, we integrate the P(z) distribution in from $z_{\rm{bin}} - 0.5$ to $z_{\rm{bin}} + 0.5$ and the galaxy is assigned to the $z_{\rm{bin}}$ with the largest integral. We then keep all sources assigned to $z_{\rm{bin}} \geq 6$.

\begin{figure*}
  \includegraphics[width=\textwidth]{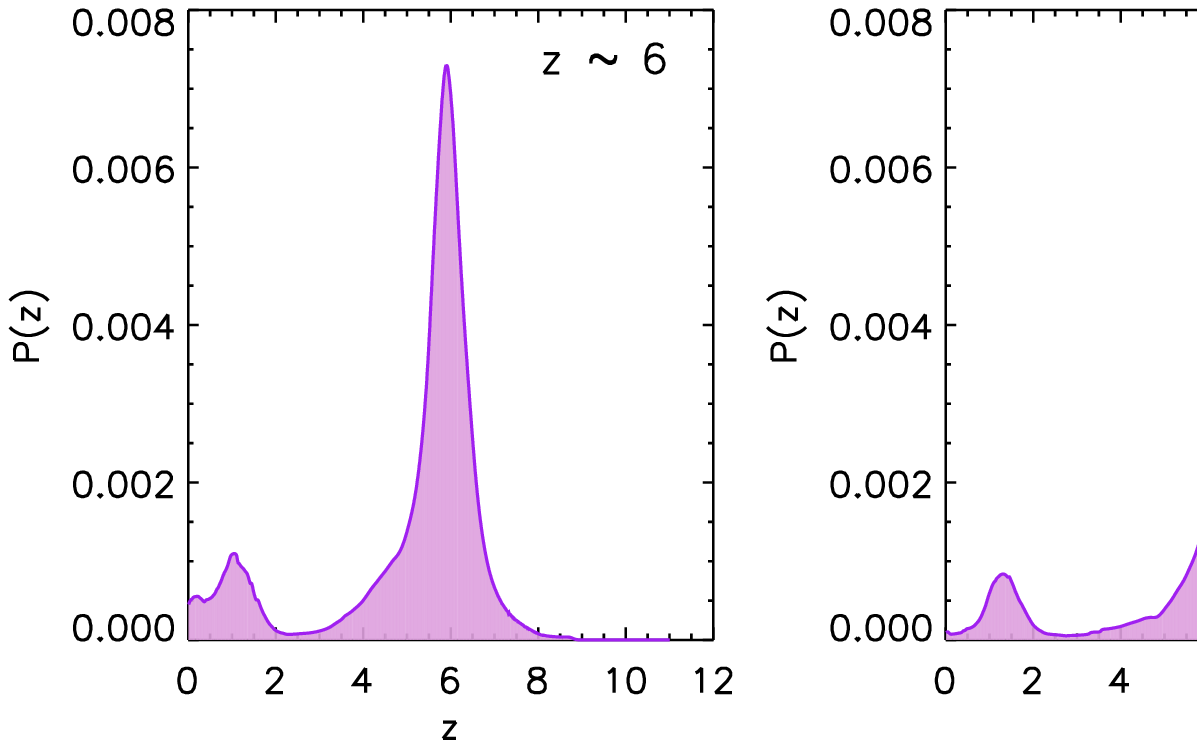}
  \caption{A stack within each redshift bin of the photometric redshift probability distributions (P($z$))for each galaxy. Our selection criteria require that the P($z$) be well-defined and strongly prefer the high-redshift solution, but most galaxies have a low-probability secondary peak at lower redshift representing the likelihood that the observed spectral break is the Balmer, rather than Lyman, break. The integral of the P(z) at $z < 3$ is 13\% at $z \sim 6$, 8.8\% at $z \sim 7$ and 5.1\% at $z \sim 8$.}
  \label{fig:pz}
\end{figure*}

\begin{figure*}
  \includegraphics[width=\textwidth]{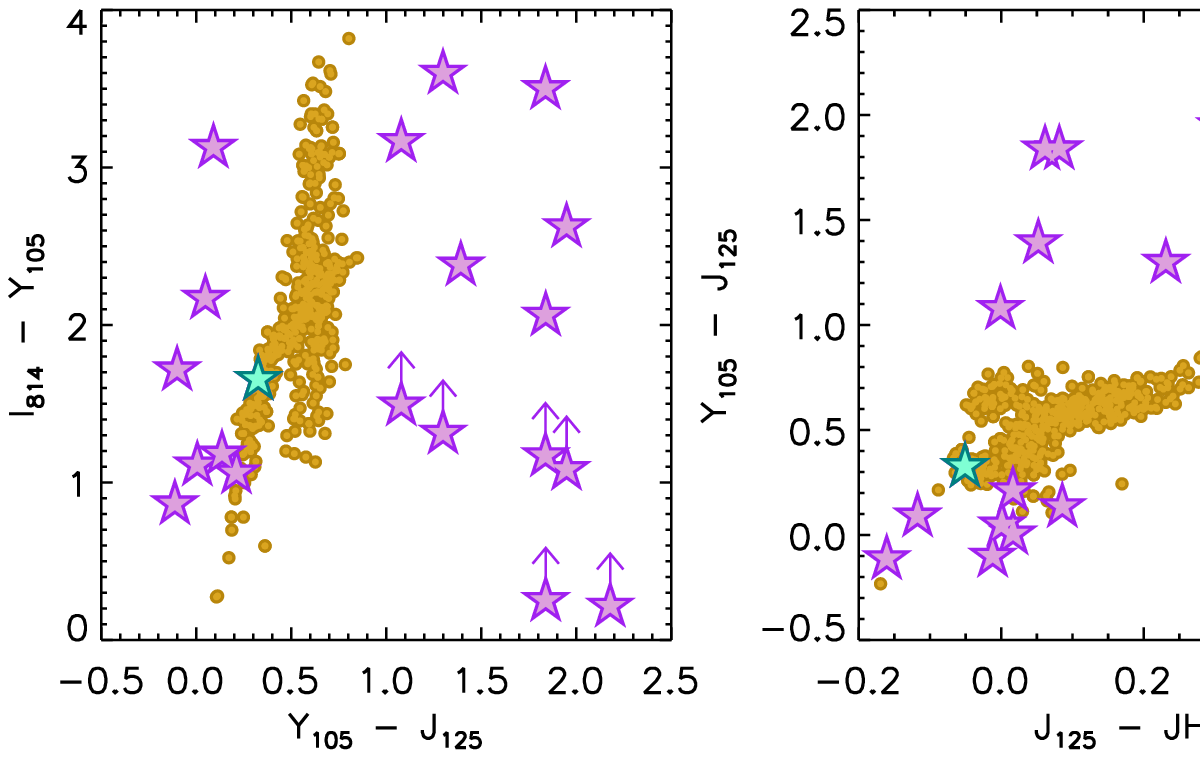}
  \caption{The colors of all $H_{160}<27$ galaxies in our sample (stars), compared to L and T brown dwarfs (yellow circles). We identify one object in our sample as being consistent with the colors of a brown dwarf (indicated by the green star). This object is also bright ($H_{160} = 24.0$) and unresolved (FWHM = 0.185''), so we exclude it from our final sample.}
  \label{fig:bd}
\end{figure*}

\subsection{Sample cleaning}

Once we have catalogs of high-redshift candidate galaxies in each cluster, we go through several steps to clean the sample of potential contaminants.

\subsubsection{Visual inspection}
\label{sec:vis}

The first step in cleaning the sample is visual inspection to remove image artifacts and spurious sources. We begin by identifying any galaxies with overlapping segmentation maps. When combining catalogs from multiple detection images, we excluded any objects within a 0.2'' matching radius, but duplicates can remain where {\sc SExtractor} defines the center of a galaxy differently in different bands. The vast majority of these duplicates are found to be large image artifacts such as stellar diffraction spikes, as these tend to be split differently in different detection images. We identify these cases through visual inspection and remove all of them from the catalogs.

After the duplicates have been removed, two members of the team (RCL and SLF) independently visually inspect all of the remaining candidates. Here, we remove any spurious sources caused by artifacts in the images, particularly those from the wavelet subtraction process, stellar diffraction spikes, and oversplit regions of brighter foreground galaxies.

In some cases, the {\sc SExtractor} segmentation map shows that objects have been incorrectly separated (either one galaxy split into multiple sources or part of a nearby source incorrectly allocated to the galaxy). For these sources, we search the other 21 catalogs and substitute the photometry from the equivalent source identified from a different detection image where it has been correctly split by {\sc SExtractor}.

\begin{figure*}
\includegraphics[width=0.5\textwidth]{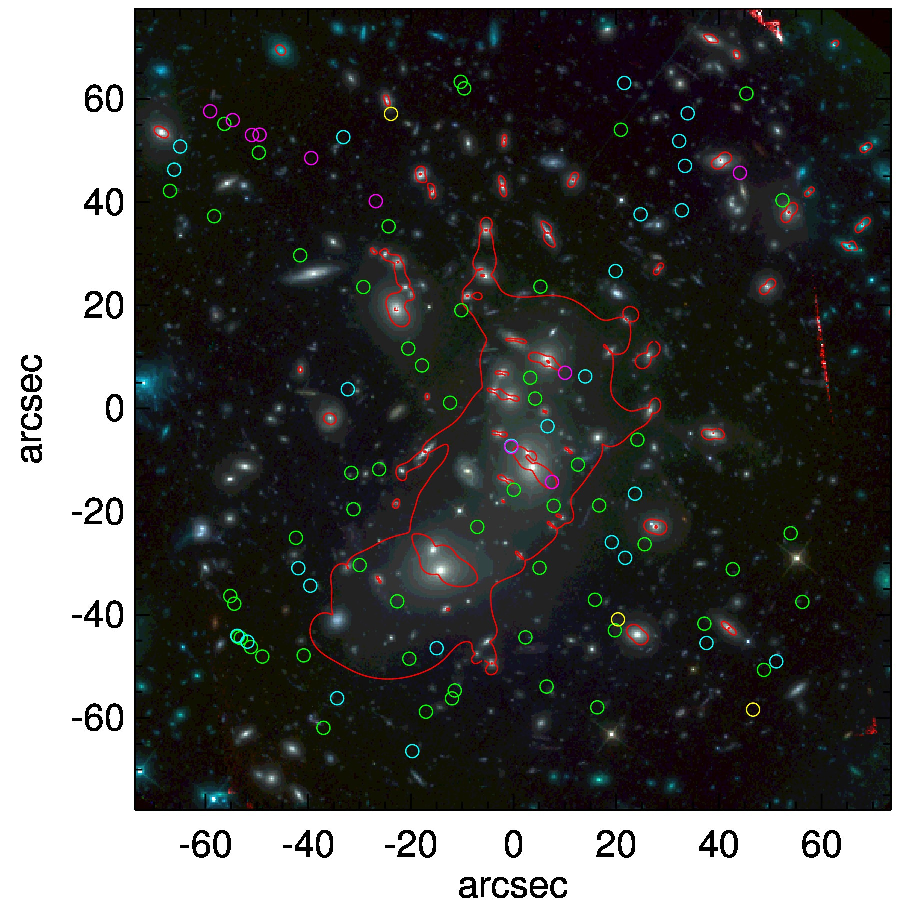} \hfill \includegraphics[width=0.5\textwidth]{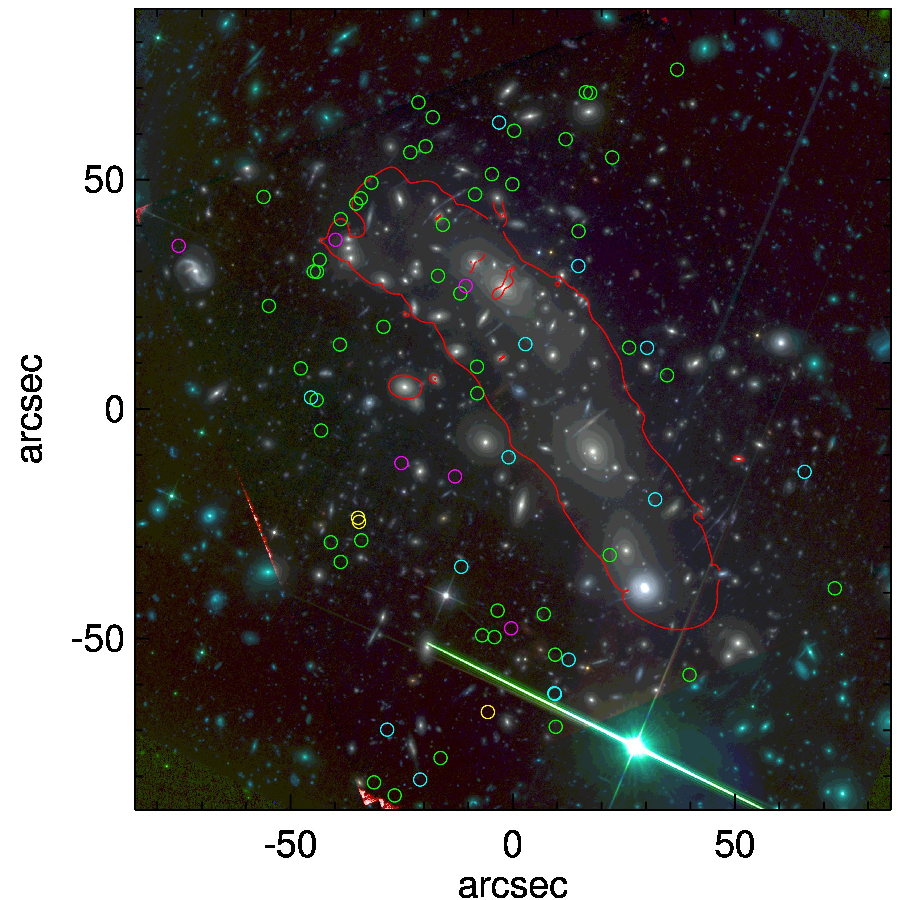}
  \caption{$V_{606}$, $I_{814}$, $J_{125}$ image of Abell 2744 (left) and MACS 0416 (right). The critical line at $z=7$ from the CATS team model \citep{2015MNRAS.452.1437J} is shown in red. The positions of the high-$z$ galaxy sample at $z = $6, 7, 8 and 9 are indicated by green, cyan, magenta and yellow circles respectively.}
  \label{fig:colim}
\end{figure*}

\begin{figure*}
  \centering
  \includegraphics[width=\textwidth]{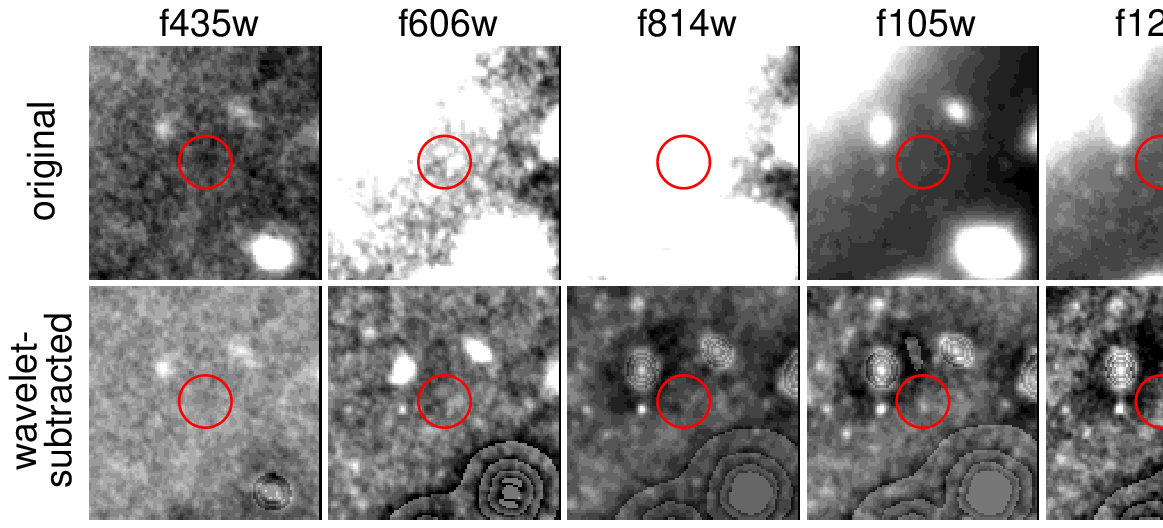}
  \includegraphics[width=\textwidth]{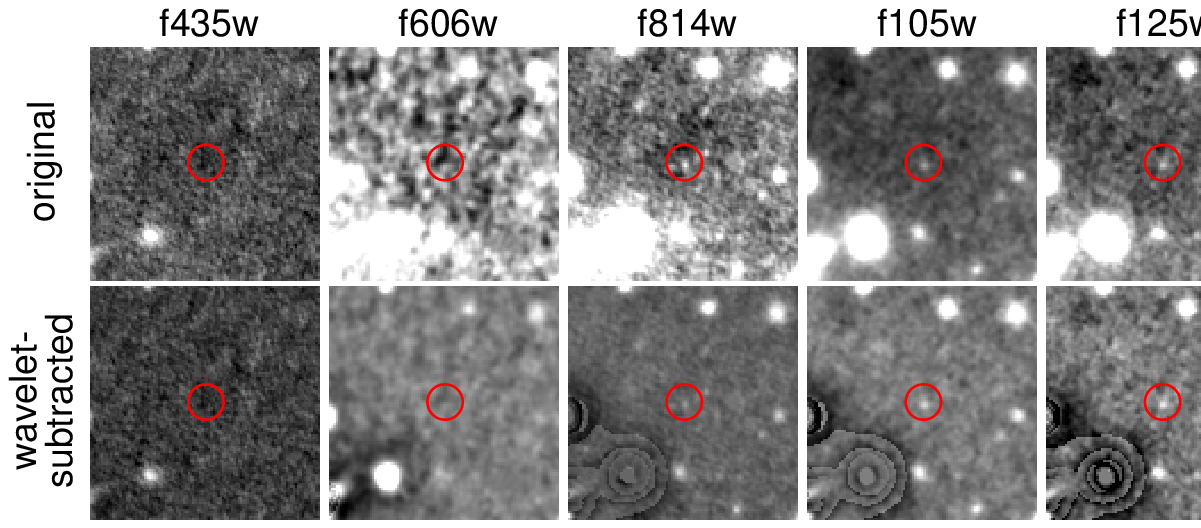}
  \caption{Postage stamp images of A2744\_z6\_2830 and M0416\_z6\_112879, the two intrinsically faintest galaxies from the $z \sim 6$ sample detected in the Abell 2744 and MACS 0416 cluster fields respectively. The circle shows a 0.4'' aperture. These galaxes are magnified by factors $\sim 110\times$ and $\sim 19\times$, giving intrinsic UV magnitudes of $M_\mathrm{UV} = -12.4$ and $-14.2$ respectively. The top row shows the original images and the lower row the wavelet-subtracted images, where removal of bright foreground galaxies close to the line of sight allows the distant galaxies to be seen more easily.}
  \label{fig:wv}
\end{figure*}

\subsubsection{Contamination}

We use the same selection method as \cite{2015ApJ...810...71F}, in which the contamination rate from faint low-redshift sources was extensively tested and found to be $<10\%$ with the adopted criteria. We show the stacked P(z) distributions in Figure \ref{fig:pz} as an indication of the likely low-redshift contamination. The total fraction of the P(z) at $z < 3$ is 13\% at $z \sim 6$, 8.8\% at $z \sim 7$ and 5.1\% at $z \sim 8$.

Another potential source of contamination noted by \citet{2015ApJ...810...71F} is that {\sc SExtractor} cannot efficiently separate stars from galaxies at faint ($J_{125} > 25$) magnitudes. A particular concern at $z \geq 6$ is contamination from L and T brown dwarfs, which can have similar colors to high-redshift galaxies. However, the colors of brown dwarfs fall on well-defined sequences and can therefore be easily separated from true galaxies.  In Figure \ref{fig:bd}, we compare the colors of all of our high-redshift galaxy candidates with $J_{125} < 27$ to the colors of brown dwarfs from the SpeX Prism Spectral Libraries\footnote{maintained by Adam Burgasser at \url{http://pono.ucsd.edu/∼adam/browndwarfs/spexprism}}. We find one source consistent with the brown dwarf colors, indicated in green in Figure \ref{fig:bd}. This source is the brightest in our sample ($H_{160} = 24.0$) and is unresolved (FWHM$ = 0.185''$), so we reclassify it as a star and remove it from the sample.

An additional potential source of contamination in this work is from globular clusters within the foreground galaxy cluster. At the redshifts of the clusters Abell 2744 and MACS 0416, we find that only very massive ($M > 10^6$) and very young ($< 1\,$Gyr) globular clusters would be detectable in these data. We consider the colors of such globular clusters and find that they are very narrowly distributed and inconsistent with the colors of galaxies in our high-redshift sample. Given the highly unlikely presence of young, high-mass globular clusters and their color distributions, we conclude that they are unlikely to be a source of contamination for our sample.

\subsection{The final sample}

The results of both visual inspections are combined to produce the final catalogs, which contain 167 galaxies; 94 in Abell 2744 and 73 in MACS 0416. The positions are shown in Figure \ref{fig:colim}, and the catalogs for the $z \sim 6$, 7 and 8 samples are listed in Tables \ref{tab:z6} -- \ref{tab:z8}. In addition, we find six galaxies at $z \sim 9$, which will be the subject of future work.

Crucially, we are able to detect galaxies close to the critical line, the line of theoretically infinite magnification close to which we benefit most from the gravitational lensing effect. Due to the critical line's location near the cluster center, there is a large number of bright foreground galaxies in this region that have hampered previous efforts to detect faint background sources. Subtraction of these foreground galaxies substantially improves the detection significance of the background source. We note that 46 of the 167 galaxies in the sample suffer from unreliable photometry in the wavelet-subtracted images, either due to partial subtraction of the source itself or due to proximity to residuals from the subtraction process. The remaining 121 (or 72\% of the sample) have their photometry measured in the wavelet-subtracted images. In order to test whether the sources with bad photometry in the wavelet-subtracted images are disproportionately faint, we compare these two samples using the KS test and find that they are consistent with being drawn from the same observed magnitude distribution with a probability of 0.992.

\subsection{Lensing Magnification}
\label{sec:mag}

\begin{table}
\begin{deluxetable*}{l c c c c c c r}
\centering
\tablecaption{Details of the lens models used \label{tab:lensmodel}}
\tablecolumns{6}
\tablewidth{0pt}
\tablehead{
  \colhead{Team} & \colhead{Version(s)} & \colhead{Version(s)} & \colhead{Area} & \colhead{Resolution} & \colhead{Light traces mass?} & \colhead{$n_{\rm{models}}$} & \colhead{References}\\
 & \colhead{pre-HFF} & \colhead{post-HFF} & & & & & }
\startdata
Bradac & 1 & 2$^a$, 3$^b$ & 3.5' $\times$ 3.5 & 0.05'' &
No & 100 & \citet{Bradac:2004tj} \\
& & & & & & & \citet{Bradac:2009bk}\\
CATS & 1 & 2, 3, 3.1 & 5.3' $\times$ 6.1' & 0.2'' & Yes, plus dark matter & 200 & \citet{2009MNRAS.395.1319J}\\
& & & & & & & \citet{2012MNRAS.426.3369J}\\
& & & & & & & \citet{2014MNRAS.443.1549J}\\
& & & & & & & \citet{2014MNRAS.444..268R}\\
& & & & & & & \citet{2015MNRAS.446.4132J}\\
glafic & 1$^a$ & 3 & 2.7' $\times$ 2.7' & 0.3'' & Yes, plus dark matter & 100 & \citet{2010PASJ...62.1017O}\\
& & & & & & & \citet{2015ApJ...799...12I}\\
& & & & & & & \citet{2016ApJ...819..114K}\\
Merten & 1 & & 25' $\times$ 25' & 8.3'' & No & 250 & \citet{2009AA...500..681M}\\
& & & & & & & \citet{2011MNRAS.417..333M}\\
Williams & 1 & 3, 3.1$^b$ & 2.3' $\times$ 2.3' & 0.3'' & No & 30 & \citet{2006MNRAS.367.1209L} \\
& & & & & & & \citet{2014MNRAS.443.1549J}\\
& & & & & & & \citet{2014MNRAS.439.2651M} \\
& & & & & & & \citet{2015ApJ...800...38G}\\
Sharon & 2 & 3 & 3.4' $\times$ 3.4' & 0.03'' &
Yes, plus dark matter &  100 & \citet{2007NJPh....9..447J} \\
& & & & & & & \citet{2014ApJ...797...48J}\\
Zitrin-NFW & 1 & 3 & 3.0' $\times$ 3.0' & 0.06'' & Yes, plus dark matter & 100 & \citet{2009MNRAS.396.1985Z} \\
& & & & & & & \citet{2013ApJ...762L..30Z}\\
Zitrin-LTM & 1 & 3 & 3.0' $\times$ 3.0' & 0.06'' & Yes & 100 & \citet{2009MNRAS.396.1985Z}\\
& & & & & & & \citet{2013ApJ...762L..30Z}
\enddata
\tablecomments{$n_{\rm{models}}$ is the number of alternative models provided by each team in addition to their best-fit model.\\
  $^a$ A2744 only\\
$^b$ M0416 only}
\end{deluxetable*}
\end{table}

Magnification estimates are highly sensitive to substructure within the cluster, so for this work we make use of the full range of possible lens models produced for the Frontier Fields by seven independent teams who use different assumptions and methodologies. The details of the models and the products are available on \href{http://www.stsci.edu/hst/campaigns/frontier-fields/Lensing-Models}{the MAST archive website}, and we summarize the characteristics of the different models in Table \ref{tab:lensmodel}. The primary difference is that some models assume that the mass in the cluster is traced by the luminous galaxies, while others operate without that assumption. The light-traces-mass assumption has been well tested and proven to improve the accuracy of lens models, but at the expense of flexibility \citep{2010MNRAS.408.1916Z,Zitrin:2011et}; models that do not assume that light traces mass allow a wider range of possible mass distributions. Both of the clusters considered here are undergoing mergers, so we might expect more deviations from light tracing mass than usual \citep{2015ApJ...811...29W}.

The magnification, $\mu$,  of a galaxy depends on the convergence ($\kappa$) and shear ($\gamma$) as

\begin{equation}
\mu = \frac{1}{\left( 1 - \kappa \right)^2 - \gamma^2}
\label{eq:mag}
\end{equation}

Both $\kappa$ and $\gamma$ scale with the distance ratio $D_{LS}/D_{S}$, where $D_{LS}$ is the angular diameter distance between the source and the lensing cluster, and $D_S$ is the angular diameter distance between the observer and the source.

The differential magnification across a galaxy image can be a strong effect close to the critical line where the magnification gradient is high. In order to account for this, we use a flux-weighted magnification calculated from the $H_{160}$-band image for each galaxy, using the best-fit normalized $\kappa$ and $\gamma$ maps provided by each lens modelling team.

In addition to their best-fit maps, each lens modelling team has made available a range of $\kappa$ and $\gamma$ maps from alternate models to enable the estimation of errors. While the redshifts of the clusters are well known, the redshifts of the background sources are subject to the errors on their photometric redshifts. When calculating errors on the estimated magnification factors, we must therefore incorporate the errors associated with $\kappa$, $\gamma$ and $D_{LS}/D_S$. We estimate errors on $\kappa$ and $\gamma$ by using the full range of lens models provided by each team. The error on $D_{LS}/D_S$ is estimated from the 68\% uncertainty range on the photometric redshift of the source (we assume that the error on the distance to the cluster is negligible).

To estimate the uncertainty on the overall magnification factors, we use a bootstrap method. We create 10,000 realizations of the magnification of each galaxy based on randomly drawn values of $\kappa$ and $\gamma$ from within the range of alternate lens models provided by each team, and a randomly selected redshift from within the probability distribution for the galaxy. We find that the uncertainty on the photometric redshift dominates over the lens model uncertainties except in cases where the magnification is high ($\mu \gtrsim 10$), where the precise position of the critical line can cause significant differences in the magnification.

In Tables \ref{tab:z6} - \ref{tab:z8} we list the median magnification across the full range of models that were updated after the Frontier Fields data were obtained (see Table \ref{tab:lensmodel} for the model version numbers). The uncertainties listed encompass the central 68\% interval over all of the models, in order to separate the model uncertainty from the photometric redshift. We use this median value for our fiducial luminosity function as it excludes the extreme magnifications which are likely unphysical. However, in order to examine the systematic uncertainties from lens modelling, we also discuss the effects of using each lens model individually in \S\ref{sec:lenserr}.

In Figure \ref{fig:wv}, we show images of the two intrinsically faintest sources ($M_{\mathrm{UV}} = -12.4$ and $-14.2$) in the sample, both before and after wavelet subtraction.

\subsection{Comparison to other samples}
\label{sec:samplecomp}

\begin{figure*}
  \centering
  \includegraphics[width=0.45\textwidth]{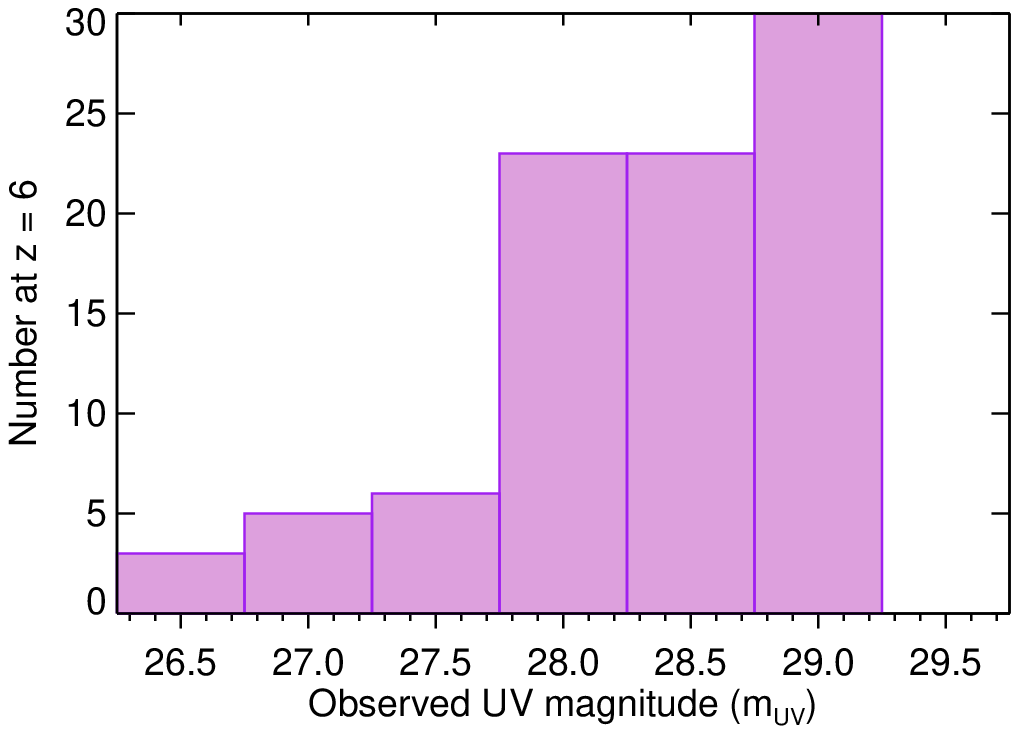} \includegraphics[width=0.45\textwidth]{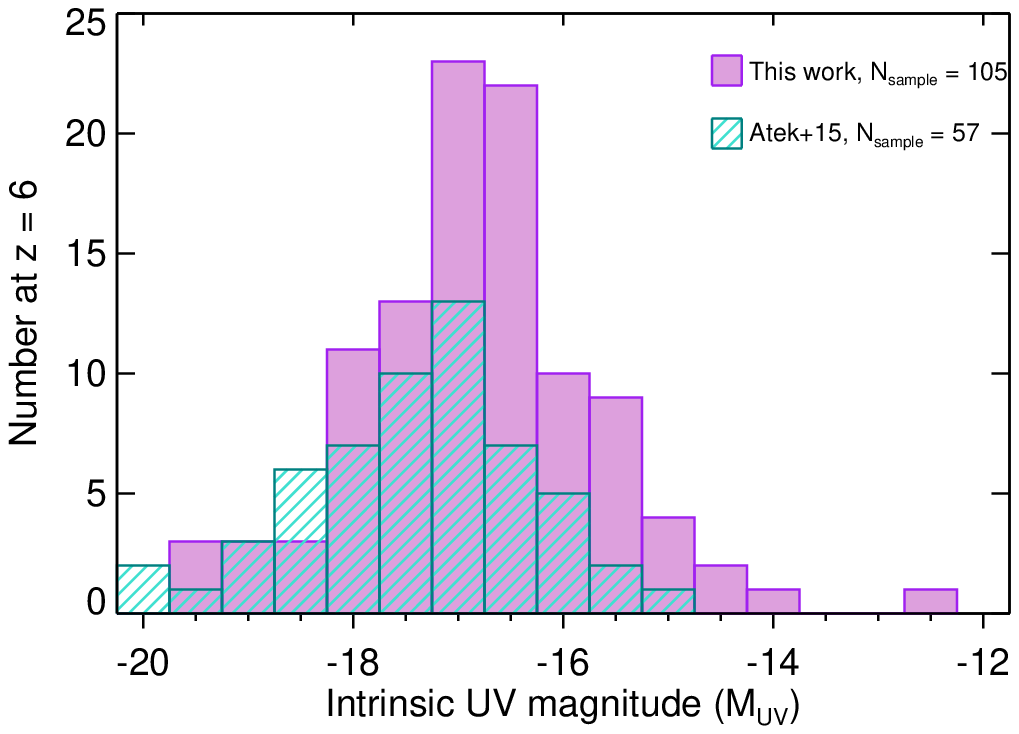} \includegraphics[width=0.45\textwidth]{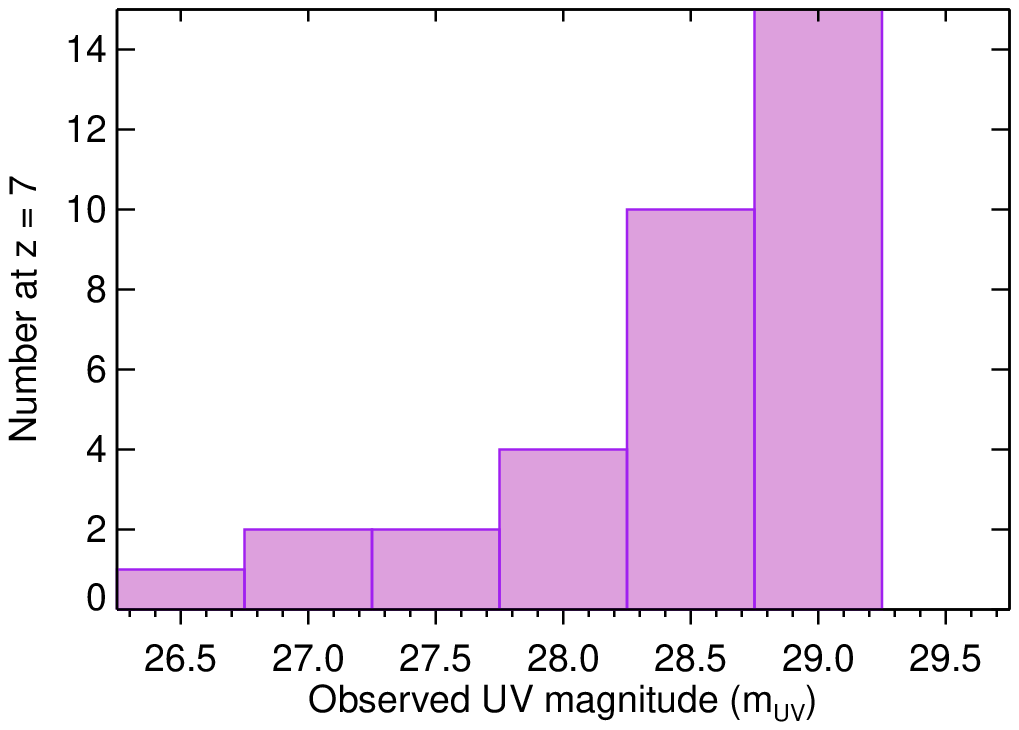} \includegraphics[width=0.45\textwidth]{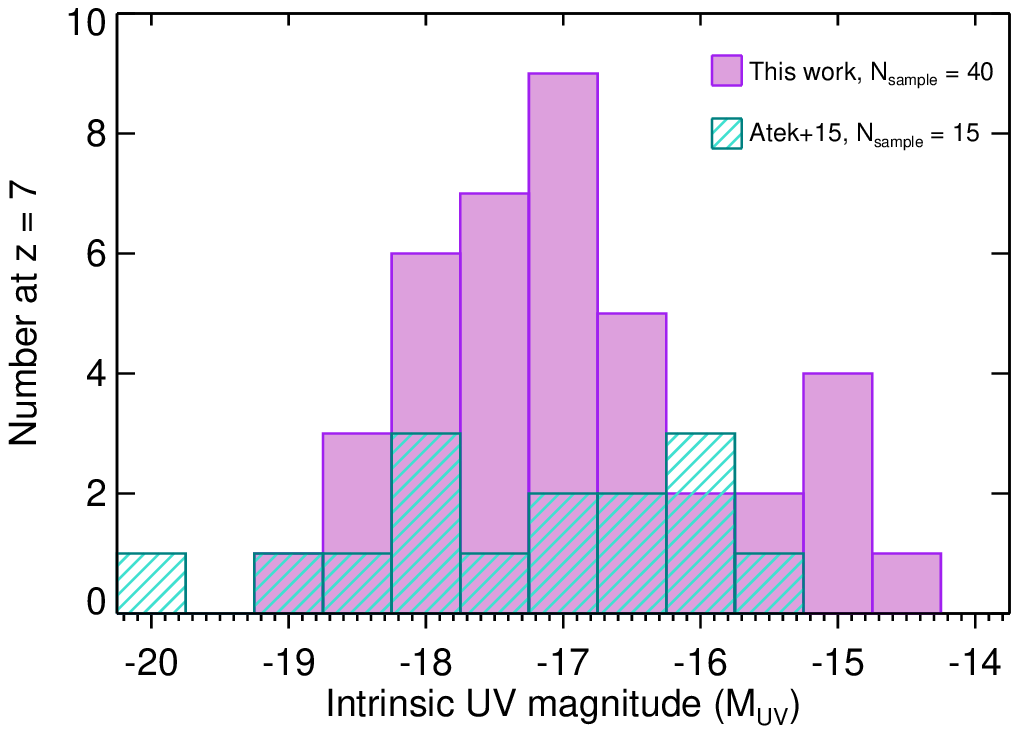} \includegraphics[width=0.45\textwidth]{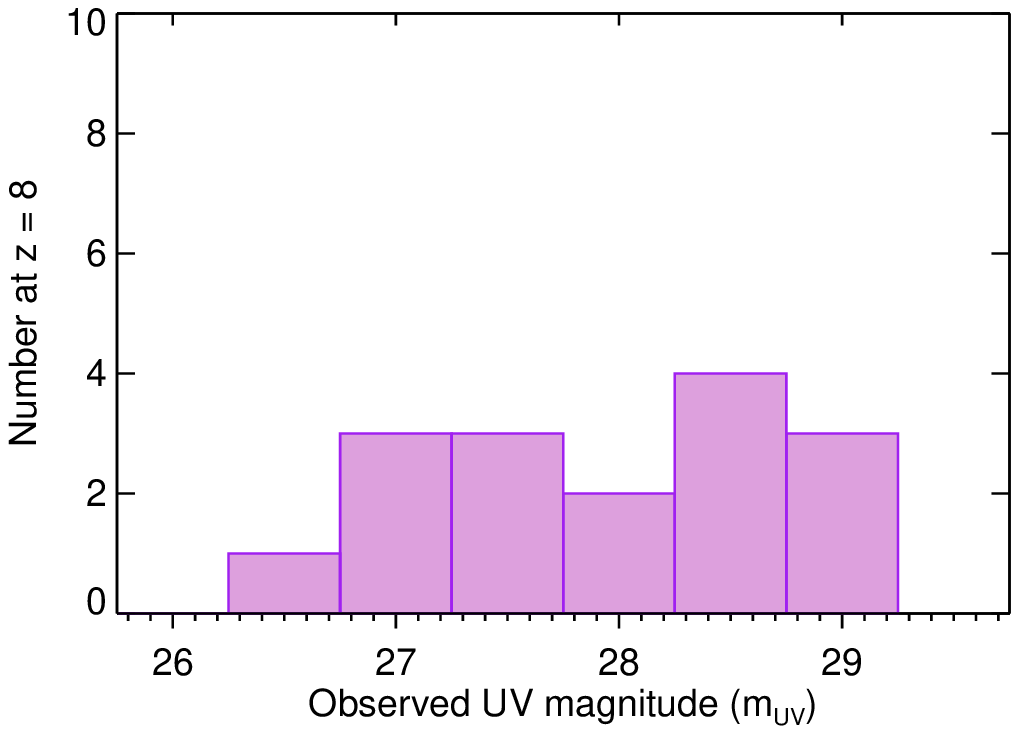}
  \includegraphics[width=0.45\textwidth]{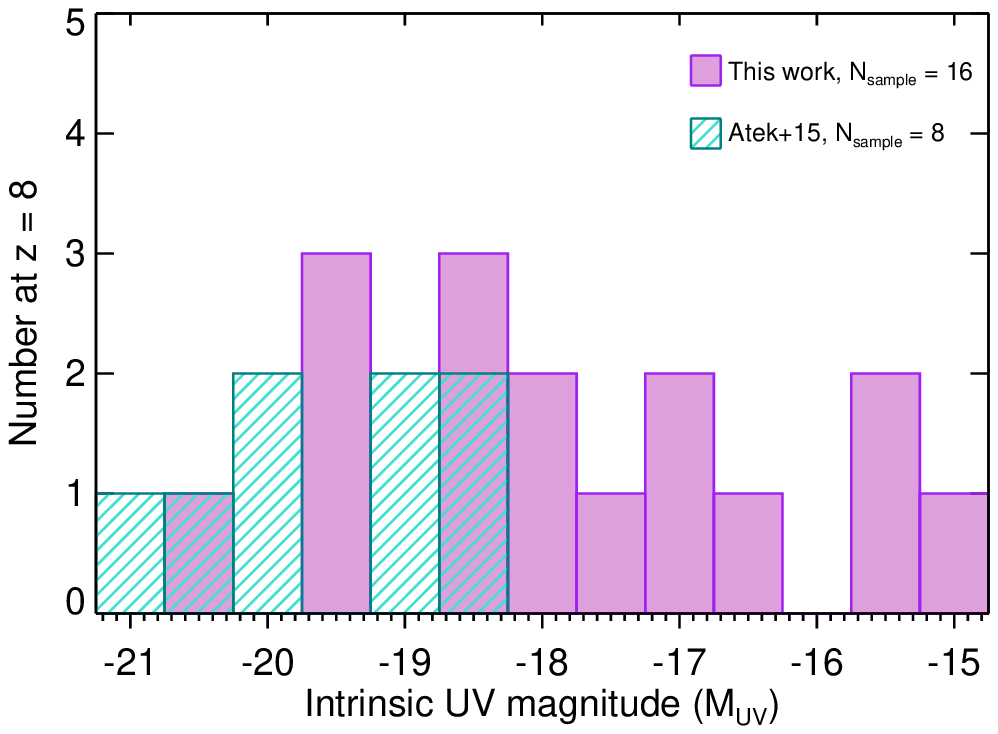}
\caption{\emph{Left:} Observed apparent magnitudes at rest-frame 1500\AA~ of the sample. \emph{Right:} Comparison between the sample in this work and that of \citet{2015ApJ...814...69AA} in the same fields. Due to the use of the wavelet decomposition technique, we are able to recover $\sim 2\times$ more galaxies in these clusters, and find galaxies that are systematically more magnified and hence intrinsically fainter, by 3 magnitudes at $z\sim 6$ and 8, and 1 magnitude at $z \sim 7$.}
\label{fig:samplecomp}
\end{figure*}

The most comprehensive previous study of the $z = 6-8$ luminosity function from the Frontier Fields to date by \citet{2015ApJ...814...69AA} contains 253 galaxies, of which 88 are selected in the two fields Abell 2744 and MACS 0416 (a further 74 are from these clusters' parallel fields; objects in these fields are included with our unlensed comparison sample from \citet{2015ApJ...810...71F}). They combine their $z \sim 6$ and $z \sim 7$ samples into one $z \sim 7$ luminosity function, so in order to compare this sample to ours, we use their photometric redshifts (supplied via private communication) to split the 88 galaxies from the first two cluster fields into the same redshift bins ($z \sim 6, 7, 8$) used in this work. The results are shown in Figure \ref{fig:samplecomp}; it is likely that some galaxies scatter between bins due to different magnification estimates, but we find $\sim 2\times$ more galaxies and probe systematically fainter than this previous work.

To test whether the larger sample size is caused by more efficient detection or differences in the selection critera, we reselect our high-redshift catalog using the wavelet subtraction method, but with the selection criteria of \citet{2015ApJ...810...71F}. Briefly, these comprise color-color cuts to select a strong break, a non-detection in the optical bands and a signal-to-noise cut in the infrared. With these criteria - and after carrying out visual inspection to remove spurious sources - we find a total of 204 galaxies between the two cluster fields; of these, 119 are in Abell 2744 and 85 in MACS 0416. There are 7 galaxies in our sample that would not be included using these criteria (all brighter than $M_{\mathrm{UV}} = -15$), leaving 44 new sources that are not in our sample. In all cases, these new sources do not meet the photometric redshift quality criteria we use in our sample selection. This implies that our selection criteria are comparatively strict, but that we find more sources overall due to the subtraction of the ICL.

\section{The Luminosity Function}
\label{sec:lf}

The rest-frame UV luminosity function provides useful insight into galaxy evolution and a direct comparison to models. Observations of the luminosity function at a wide range of redshifts have shown that the density of galaxies $\phi\left(L\right)$ is characterised by a power law with an exponential decline at the bright end, parameterised by a Schechter function \citep{1976ApJ...203..297S} of the form

\begin{equation}
  \phi\left( L \right) = \phi^{\ast}\left(\frac{L}{L^{\ast}}\right)^\alpha \exp\left( -\frac{L}{L^{\ast}}\right).
\end{equation}

The Schechter function can also be expressed in terms of magnitudes as

\begin{equation}
  \phi\left(M\right) = 0.4 \ln \left(10\right)\phi^{\ast}10^{0.4\left(M-M^{\ast}\right)\left(\alpha+1\right)}e^{-10^{-0.4\left(M-M^{\ast}\right)}}
  \label{eq:schechter}
\end{equation}

and is described by three parameters: the characteristic luminosity $L^{\ast}$ (or magnitude $M^{\ast}$), the characteristic number density $\phi^{\ast}$ and the faint-end slope $\alpha$.

In the redshift range under consideration, the integral of the UV luminosity function provides a constraint on the contribution of galaxies to reionization. Previous work has shown that there could be sufficient ionizing radiation originating in galaxies at $z > 6$ to power reionization \textit{if} the luminosity function continues unbroken to $M_{\rm{UV}} = -13$ \citep{2015ApJ...810...71F,2015ApJ...802L..19R}, well below the \textit{HST} detection limit of $M_{\rm{UV}} = -17$ or previous lensed studies that extend to $M_{\rm{UV}} = -15.25$ \citep{2015ApJ...814...69AA}. However, results from simulations indicate that the luminosity function may flatten or turn over entirely at the faint end due to the quenching of star formation in low-mass halos. In some simulations, this results in far fewer faint galaxies than are required to sustain reionization \citep{2013ApJ...766...94J,2015ApJ...807L..12O,2015arXiv151200563L}, though others show a break below the limit required for reionization \citep[e.g.][]{2011MNRAS.410.1703F,2016arXiv160401314Y,2016arXiv160307729G}. Low-redshift observations also imply that there must be a break in the luminosity function at $M_{\rm{UV}} \sim -13$ at $z \sim 7$ in order to avoid overproducing dwarf galaxies in the Local Group \citep{2014MNRAS.443L..44B,2015MNRAS.453.1503B}.

It is therefore particularly crucial to constrain the faint-end slope $\alpha$ and to confirm that the luminosity function continues to rise at the faint end.

\subsection{Completeness Simulations}
\label{sec:complete}

In order to compute the effective volume for the luminosity function, we carry out simulations to determine the recovery fraction of galaxies as a function of magnitude, position, size and color using 100,00 CPU hours provided by the Texas Advanced Computing Center (TACC). The method is similar to that employed in Section \ref{sec:wvtest} to determine the reliability of the wavelet decomposition process, but we now allow a range of intrinsic sizes and colors for the fake galaxies, incorporate the effect of lensing on their shapes, attempt to recover them using the full range of 22 detection images as described in Section \ref{sec:catalog}, and finally carry out photo-$z$ fitting on the recovered galaxies in the same manner as for the real galaxies and incorporate the effects of our photo-$z$ selection.

The first step is to create the fake galaxies. For computational efficiency, we need to simulate multiple galaxies in each iteration, without adding artificial confusion. We therefore carry out preliminary tests to determine the optimal number of galaxies to add to each image. The results are unchanged when up to 500 galaxies are added at a time, so we conservatively opt for 250. For the sizes, we use a normal distribution of half-light radii with a peak at 0.5\,kpc (1.5 pixels) and hard cutoffs at 0 and 5\,kpc ($\sim 14.5$ pixels at $z \sim 6$; \citealt{2015ApJ...804..103K}; \citealt{Shibuya:2015uv}). We choose Sersic indices ($n$) from a log-normal distribution between $1 < n < 5$ with a peak at $n \sim 1.5$ (for disc-like morphologies). The axial ratio is also log-normal with a peak at 1.8, and the position angle on the sky is selected from a uniform random distribution between 0 and 360$^{\circ}$. To obtain a range of colors, we define ranges of redshifts, ages, E(B-V) and metallicities. The redshift is selected from a uniform distribution between $z = 5 - 11$, extending above and below the redshift range of interest. In each realization of the simulation, we use the same redshift for all sources to simplify the transformation to the image plane. We assign a dust attenuation factor E(B-V) from a normal distribution centered on 0.1 with $\sigma = 0.15$, restricted to the range 0 -- 0.5. The age is a lognormal distribution peaked at 7\,Myr and limited to less than the age of the Universe at the specified redshift. We also use a lognormal distribution in metallicity peaked at 0.2\,Z$_{\odot}$. These parameters together combine to provide a distribution of UV slopes $\beta$, designed to match the distribution we expect in galaxies at these redshifts \citep{2012ApJ...756..164F,2014ApJ...793..115B}. Given these distributions, we use the stellar population synthesis models of \citet{2003MNRAS.344.1000B} to calculate the colors of the galaxies. We then select the $H$-band magnitudes from a uniform distribution between 22 and 38, and use the colors derived from the stellar population models to obtain the magnitudes in the other six filters.

We use the magnitudes, sizes and shapes selected above to generate a $101\times 101$ pixel postage stamp image of each source with {\sc galfit}. We assign positions to the fake sources uniformly in the image plane, and then trace this position back to the source plane using the lens model. This method ensures complete coverage of the image plane, whereas uniformly sampling the source plane causes the more magnified regions to be undersampled. The fake galaxy images are each added to a blank image of the source plane of the field at the appropriate positions, and then lensed back to the image plane using {\sc lenstool} \citep{1993PhDT.......189K,2007NJPh....9..447J,2009MNRAS.395.1319J} with the latest lens models \citep{2015MNRAS.452.1437J,2015MNRAS.446.4132J}. This enables us to account for some lensing shear in our recovery fractions. Due to the computational time required for these simulations, we are limited to using a single lens model. Since the completeness is calculated in the image plane the only uncertainty this introduces is in the difference between lensing shears in different models. This is not expected to be a dominant effect, but is a limitation of these time-intensive simulations. Once the galaxies have been lensed, we convolve the image with the $H_{160}$-band PSF and use {\sc SExtractor} to recover them and measure the image-plane half-light radii and magnitudes. These, together with all of the input parameters above and the positions in the image and source planes, are saved into an input array. We then add the images consisting of fake, lensed galaxies to the real images in each filter. We then carry out the wavelet decomposition procedure described in 
Section \ref{sec:wv}, construct the full range of 22 detection images listed in Section \ref{sec:catalog} and construct new catalogs using the same method employed for the real data. From these catalogs, we match to the input fake source catalog using a 0.2'' matching radius. We then measure the photometric redshifts of these recovered sources using {\sc EAZY} in the same manner employed for the real catalog, and apply the same cuts as described in Section \ref{sec:photoz} to determine which galaxies we would have selected in our high-redshift sample.

We repeat the above process 2,000 times, resulting in 500,000 fake galaxies in each cluster. This number is required because the completeness is highly position-dependent in strong lensing fields. We then collate the results and calculate the recovery fraction as a function of position and observed magnitude.

\begin{figure*}
  \includegraphics[width=\textwidth]{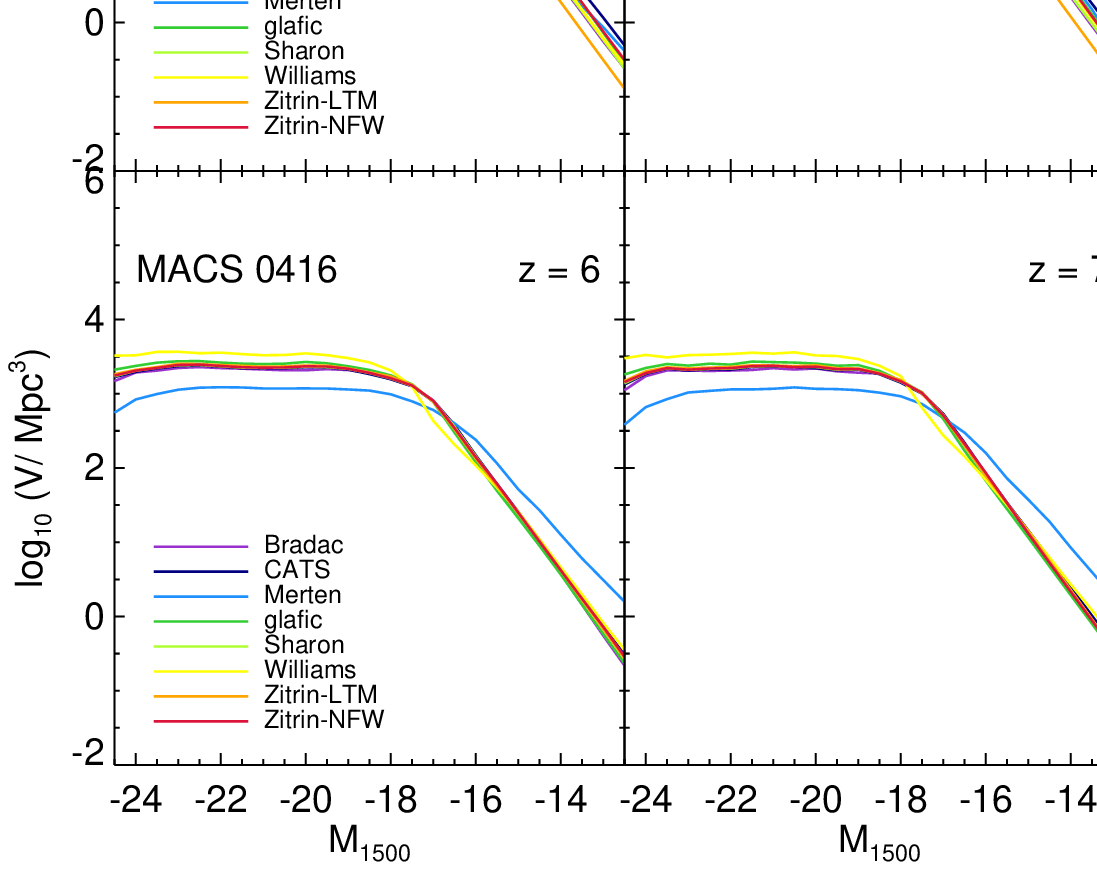}
  \caption{Effective volume as a function of intrinsic rest-frame UV magnitude for each cluster and lens model, incorporating completeness calculated as described in the text.}
  \label{fig:vol}
\end{figure*}

The effective volume, $V_m$, for a given intrinsic magnitude $m$ is given by

\begin{equation}
  V_{m} = \sum\limits_{i=0}^N \frac{v}{\mu_i} f_{i - \mu_{i,\rm{mag}}}
\end{equation}

where $N$ is the number of pixels in the image, $v$ is the volume of one unlensed pixel, $\mu_i$ is the linear magnification factor of pixel $i$, $f_{i - \mu_{i,\rm{mag}}}$ is the completeness fraction in pixel $i$ at image-plane magnitude $i - \mu_{i,\rm{mag}}$, and $\mu_{i,\rm{mag}}$ is the magnification in pixel $i$ expressed in magnitudes. The resulting total volume as a function of magnitude for each cluster, lens model, and redshift bin is shown in Figure \ref{fig:vol}. Note that due to computational runtime we do not use the shear from different lens models, but we do account for the different magnifications when calculating the effective volumes.

Calculating the volume in the image plane in this way means that regions which are multiply imaged are counted multiple times, with the appropriate recovery fraction for each of the multiple images. By double-counting both the multiply imaged galaxies and their respective volumes, we can eliminate errors in the luminosity function due to misidentified multiple images.

\subsection{The UV Luminosity Function}

\begin{figure*}
\begin{center}
\includegraphics[width=0.49\textwidth]{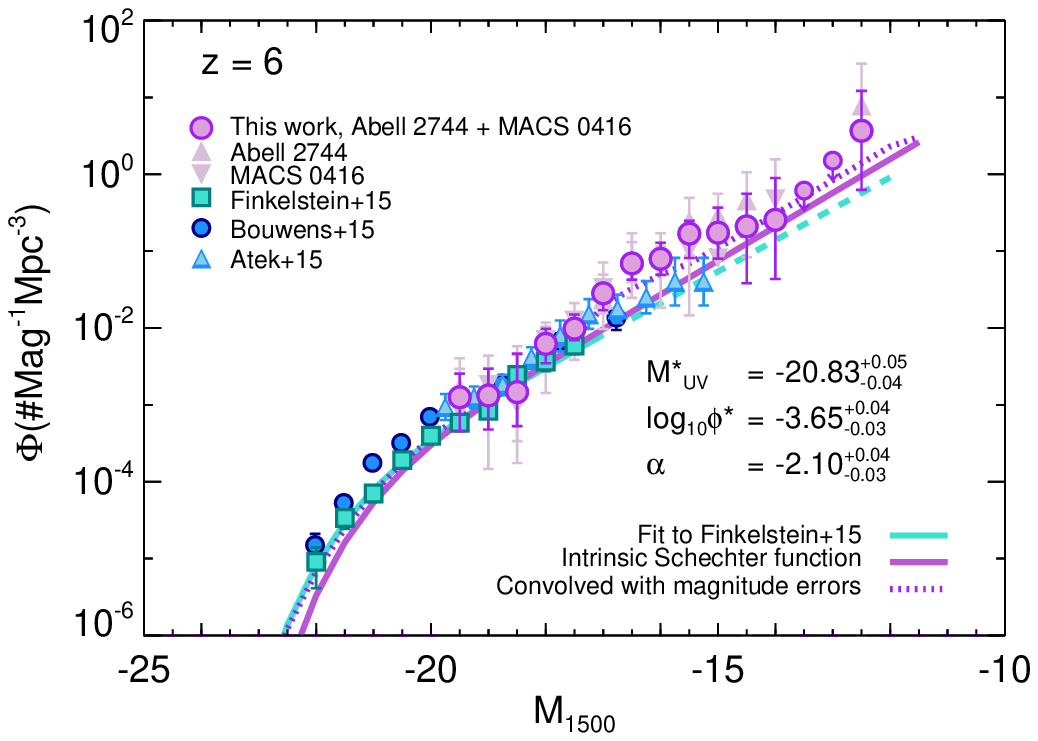}
\includegraphics[width=0.49\textwidth]{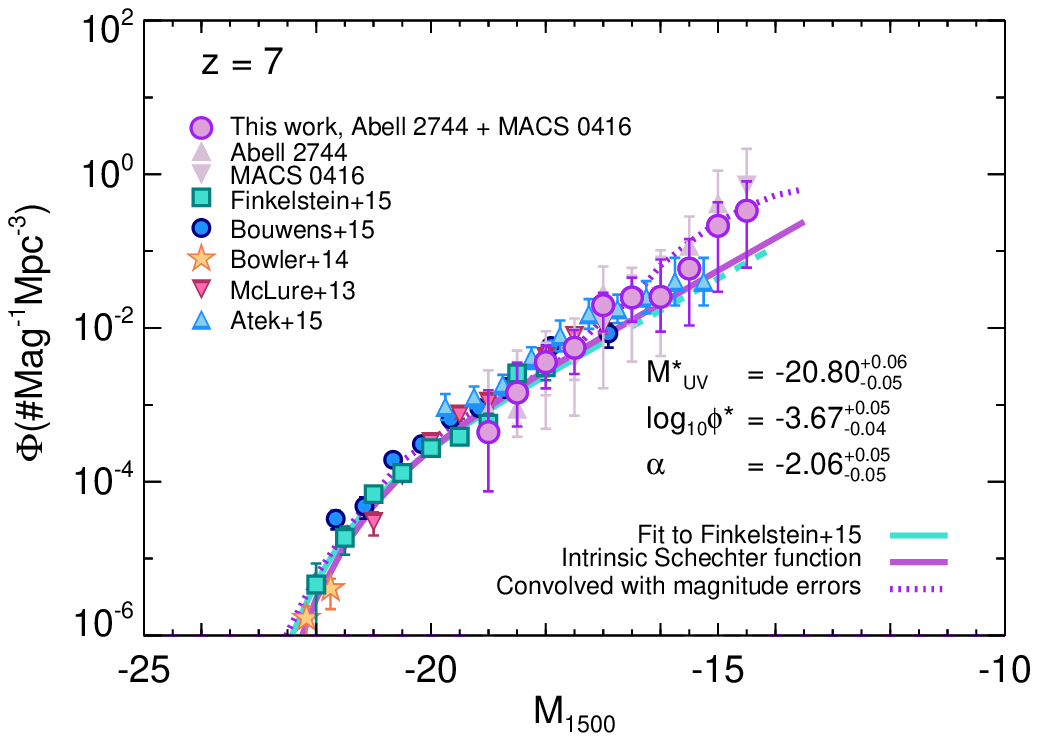}
\includegraphics[width=0.49\textwidth]{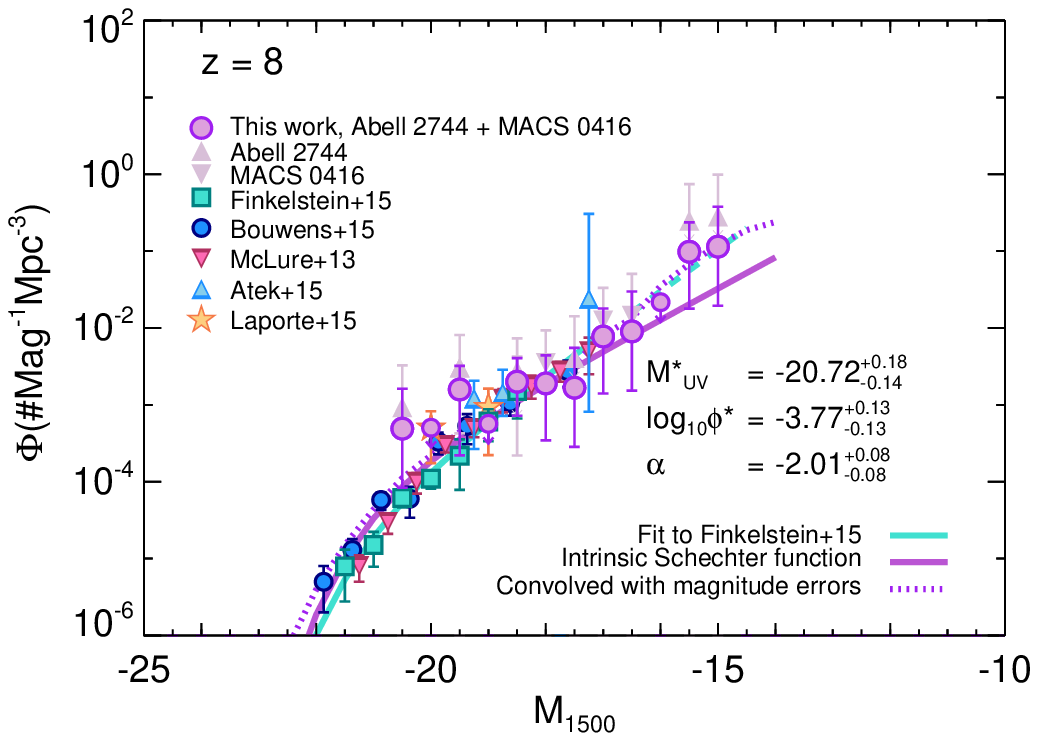}
\end{center}
  \caption{Rest-frame UV luminosity functions at $z = 6-8$ from the Hubble Frontier Fields samples. The solid green line is the fit to the CANDELS data from \citet{2015ApJ...810...71F}, and the green dashed line extends this fit to fainter magnitudes. Error bars are computed from the Poisson error based on the number of galaxies in each bin and the uncertainty in the number of galaxies due to the magnitude errors; cosmic variance and uncertainties in magnitude and redshift are also accounted for when fitting the Schechter parameters. The solid purple line is the intrinsic Schechter function fit to the combined HFF and CANDELS data, and the dotted line shows the intrinsic Schechter function after convolving with the magnification errors to account for Eddington bias. The best-fitting Schechter function parameters are shown with the combined statistical and systematic uncertainties; these are given separately in Table \ref{tab:lfpars}.}
  \label{fig:lf}
\end{figure*}

Using the galaxy sample from Section \ref{sec:catalog} and effective volumes based on the completeness simulations described in Section \ref{sec:complete}, we can produce the luminosity functions at $6 < z < 8$. The resulting luminosity functions are shown in Figure \ref{fig:lf}. The error bars in Figure \ref{fig:lf} are calculated from a combination of the Poisson uncertainty and the uncertainty on the intrinsic magnitudes. To estimate the uncertainty on the number of galaxies in the bin, we carry out 10,000 realizations of the luminosity function in which each galaxy has a new intrinsic magnitude based on uncertainties in its photometry, magnification (using the bootstrap method described in Section \ref{sec:mag}) and photometric redshift. The final uncertainty on the number density is the 68\% central confidence interval of the number of galaxies in each bin over all realizations. This is added in quadrature to the Poisson uncertainty, for which we use Equation 33.59 of \citet{Nakamura:2010dr} appropriate for small numbers.

In order to fit Schechter function parameters to our luminosity functions, we employ a Markov Chain Monte Carlo (MCMC) algorithm to search the parameter space. Since our data constrain only the faint end, we include unlensed data from the CANDELS fields that use the same selection method \citep{2015ApJ...810...71F} in order to add constraints at the bright end. In absolute magnitude bins of $\Delta M$ = 0.5 and for each field individually, we calculate the number of galaxies expected for the given Schechter parameters $M^{\ast}$, $\phi^{\ast}$ and $\alpha$. Within each subfield we add a randomly-generated value for cosmic variance based on a normal distribution with the width of the expected fractional cosmic variance in that field \citep{2014ApJ...796L..27R}. This is compared to the number of observed galaxies in each bin, which is perturbed on each iteration within the uncertainties derived above. It is also important to account for the systematic effect of errors in magnitudes, which can cause an ariticially steepened observed slope in the luminosity function due to Eddington bias. Since the intrinsic slope of the luminosity function is steep, large magnitude errors mean that galaxies preferentially scatter into brighter bins. This has been shown to be especially important at the faint end, where low signal-to-noise means that photometric errors are larger than at the bright end \citep{2015AA...575A..96G}. This effect becomes even more important in lensing fields where the faint end consists of galaxies with high magnifications, which also have high uncertainties.

To account for the Eddington bias in our sample, we construct a magnitude probability distribution function for each galaxy. This is derived by selecting 10,000 realizations of the magnitude of each galaxy by drawing randomly from within the photometric and magnification errors, and from the photo-$z$ probability distribution. We combine all of the magnitude distribution functions within each subfield to produce a probability distribution (PDF) function $P\left(M_i,M_j\right)$ that a galaxy with magnitude $M_i$ has magnitude $M_j$ consistent the uncertainties in its photo-$z$, photometry and magnification. These PDFs are narrow at the bright end, where the photometry is more certain and there is little or no lensing magnification, but become much broader at the faint end where all of these uncertainties are higher.

To calculate the expected luminosity function in each subfield $f$ and magnitude bin $M_i$, we have

\begin{equation}
\phi_i = \sum_{j=0}^N\phi_{j,\rm{int}}\left(1 + CV_j\right)P\left(j,i\right)
\label{eq:phiexp}
\end{equation}

where $CV_j$ is the cosmic variance estimate in magnitude bin $M_i$, drawn from a random normal distribution with the width of the estimate of fractional cosmic variance from \citet{2014ApJ...796L..27R} and $\phi_{j,\rm{int}}$ is the intrinsic Schechter function at magnitude $j$.

For each combination of Schechter parameters $M^{\ast}$, $\phi^{\ast}$ and $\alpha$, we calculate the goodness-of-fit statistic

\begin{equation}
\mathcal{C}^2\left(\phi\right) = -2\ln\mathcal{L}\left(\phi\right)
\end{equation}

where $\mathcal{L}$ is the likelihood that the number of galaxies observed in that field and magnitude bin matches the number expected according to Equation \ref{eq:phiexp}. The final goodness-of-fit $\mathcal{C}^2$ is the sum over all fields and magnitude bins at a given redshift.

We use an {\sc IDL} implementation of an affine-invariant ensemble MCMC sampler \citep{2013PASP..125..306F,2015arXiv151105558F} to search the parameter space. For each redshift, we use $10^3$ independent chains of $10^5$ steps each, computing the likelihood of the given Schechter parameters at each step. Using independent chains prevents the fit being trapped by local minima in the parameter space. To compute the final best fit, we join all of the chains together to give $10^8$ Schechter function parameters for each redshift. The best-fit parameters listed in Table \ref{tab:lfpars} are the median of this distribution, with errors covering the central 68\% of values.

It should be noted that the two faintest bins in our $z \sim 6$ luminosity function that contain data are based on just one galaxy each. These are highly magnified and have correspondingly large uncertainties in their magnifications; different lens models give a magnification factor for the faintest source of anywhere between $23\times$ (Zitrin\_LTM\_Gauss) and $150\times$ (Bradac v2, Sharon v2), with a median of $100\times$. The MCMC code used to fit the Schechter parameters should be more strongly constrained by bins containing large numbers of galaxies, but to ensure that these two bins are not skewing the results we fit new parameters at $z \sim 6$ with these galaxies entirely excluded. We obtain exactly the same value for the faint-end slope, with error bars larger by 6\%.

\begin{table}
\begin{deluxetable*}{c c c c}
  \centering
  \tablecaption{Best-fitting Schechter function values \label{tab:lfpars}}
  \tablecolumns{4}
  \tablewidth{0pt}
  \tablehead{
    \colhead{Redshift} & \colhead{$M^{\ast}$} & \colhead{$\log \phi^{\ast}$} & \colhead{$\alpha$} }
  \startdata
    6 & $-20.825^{+0.051}_{-0.040}$\,(stat.)\,$^{+0.004}_{-0.003}$\,(sys.) & $-3.647^{+0.037}_{-0.033}$\,(stat.)\,$^{+0.002}_{-0.004}$\,(sys.) & $-2.10^{+0.03}_{-0.03}$\,(stat.)\,$^{+0.02}_{-0.01}$\,(sys.) \\
    7 & $-20.802^{+0.055}_{-0.050}$\,(stat.)\,$^{+0.007}_{-0.003}$\,(sys.) & $-3.673^{+0.046}_{-0.043}$\,(stat.)\,$^{+0.001}_{-0.004}$\,(sys.) & $-2.06^{+0.04}_{-0.04}$\,(stat.)\,$^{+0.01}_{-0.01}$\,(sys.) \\
    8 & $-20.723^{+0.181}_{-0.142}$\,(stat.)\,$^{+0.014}_{-0.018}$\,(sys.) & $-3.768^{+0.129}_{-0.131}$\,(stat.)\,$^{+0.022}_{-0.004}$\,(sys.) & $-2.01^{+0.08}_{-0.08}$\,(stat.)\,$^{+0.02}_{-0.02}$\,(sys.) \\
    \enddata
    \tablecomments{Parameters are fit using an MCMC method as described in the text, with a combination of the lensed HFF sample and unlensed CANDELS, HUDF and HFF Year 1 parallel field sample from \citet{2015ApJ...810...71F}. Statistical errors are the 68\% central range of the MCMC samples, and systematic errors are based on the range of available lens models.} 
\end{deluxetable*}
\end{table}

\subsection{Effect of lens model uncertainties}
\label{sec:lenserr}

For our fiducial luminosity functions, we use the median magnification for each galaxy in the sample. Here, we consider the systematic uncertainty due to the lens modelling by using each available lens model individually and looking at the full range of results of the best-fitting Schechter parameters.

\begin{figure*}
\includegraphics[width=0.45\textwidth]{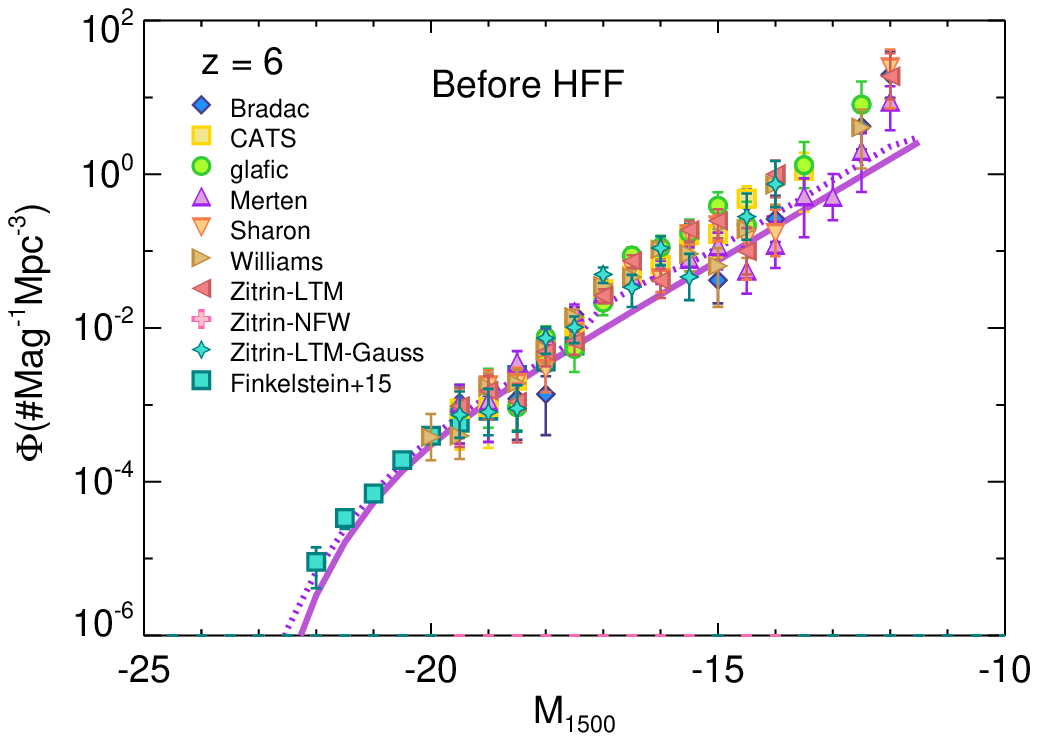} \hfill \includegraphics[width=0.45\textwidth]{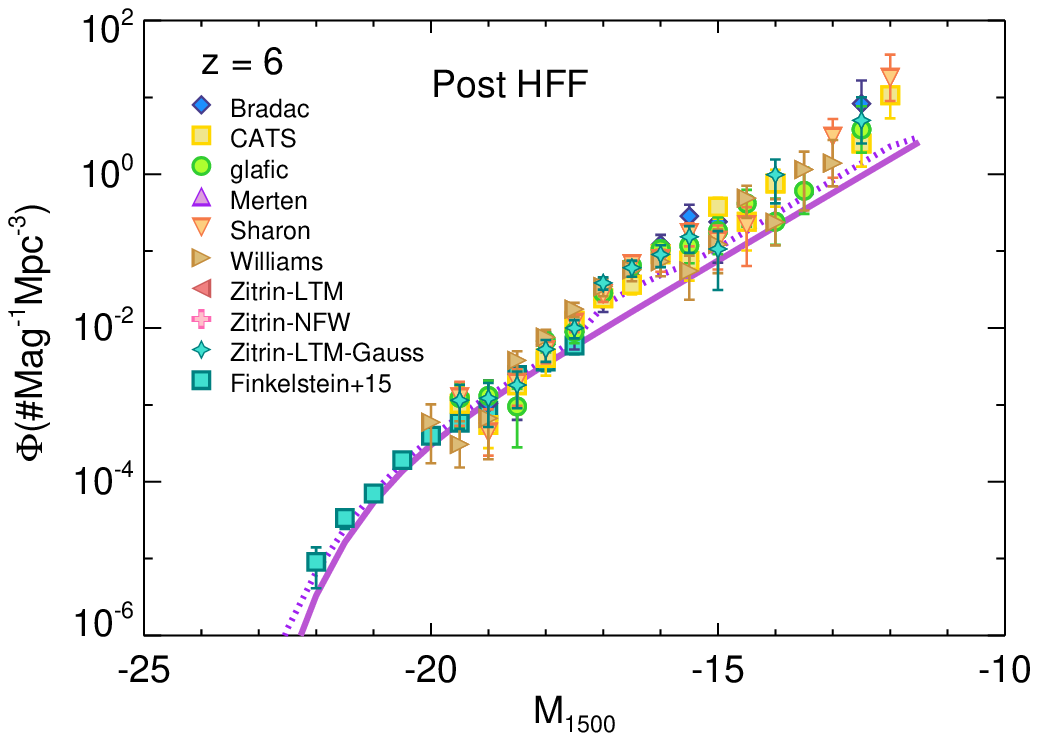} \\
\includegraphics[width=0.45\textwidth]{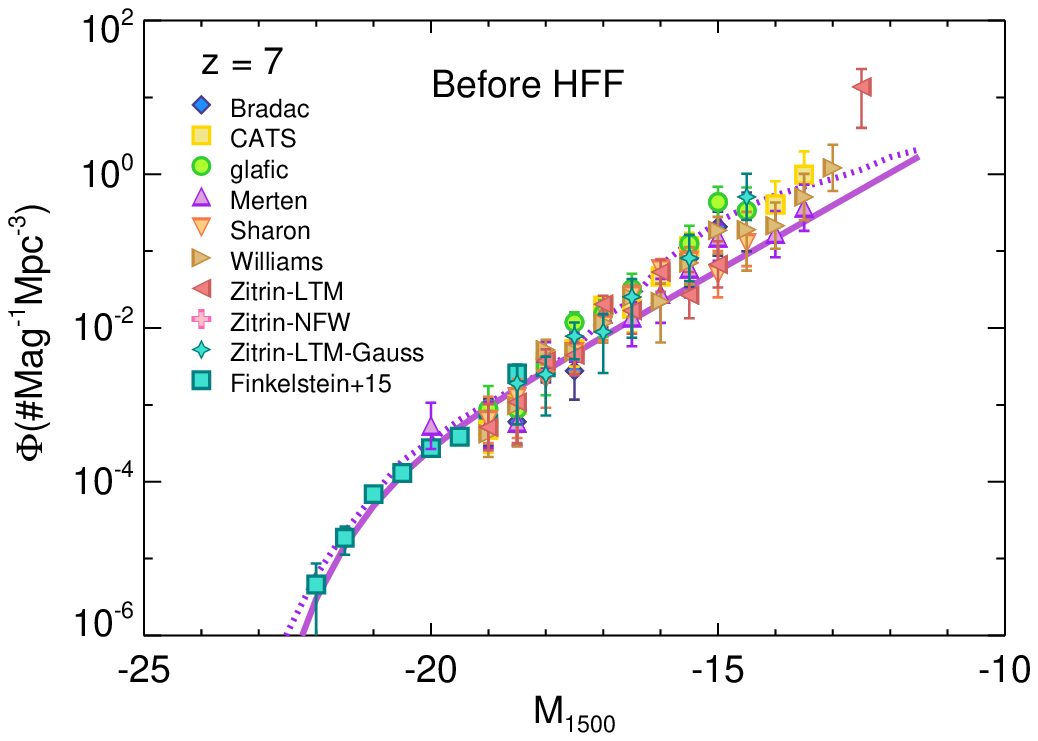} \hfill \includegraphics[width=0.45\textwidth]{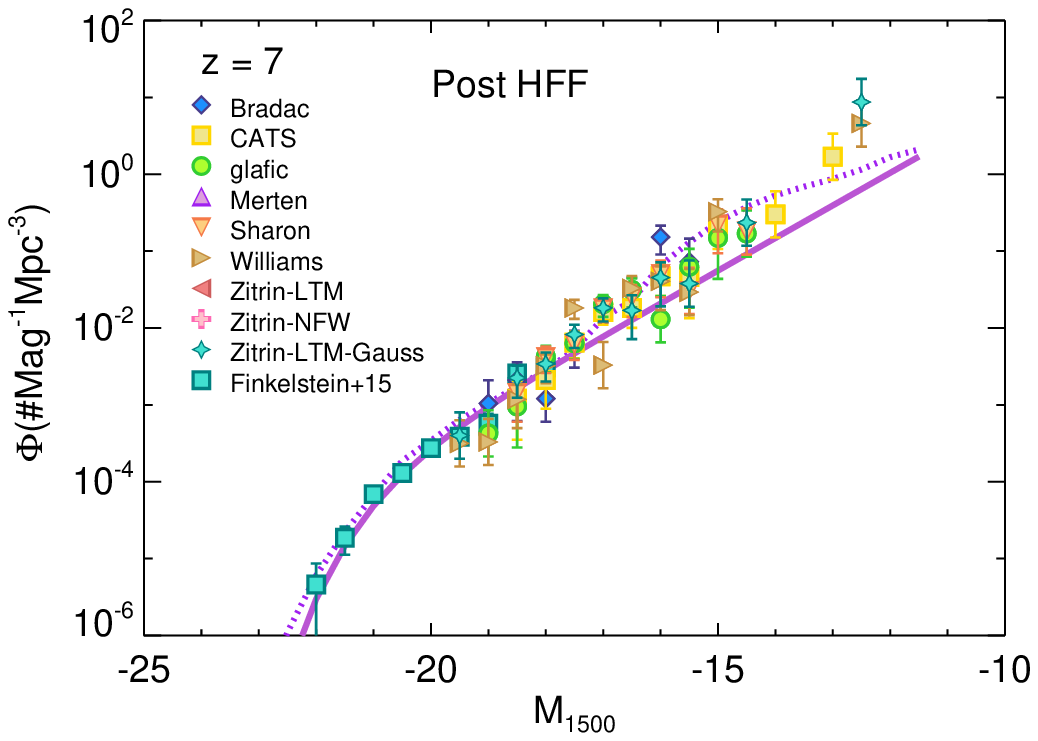} \\
\includegraphics[width=0.45\textwidth]{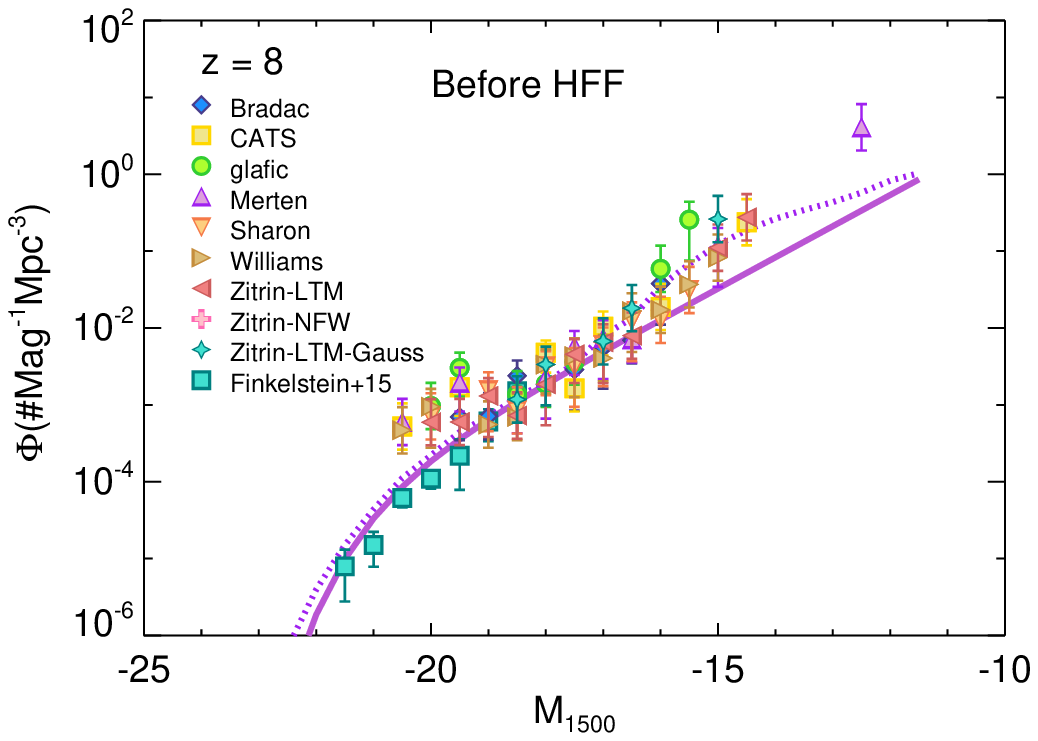} \hfill \includegraphics[width=0.45\textwidth]{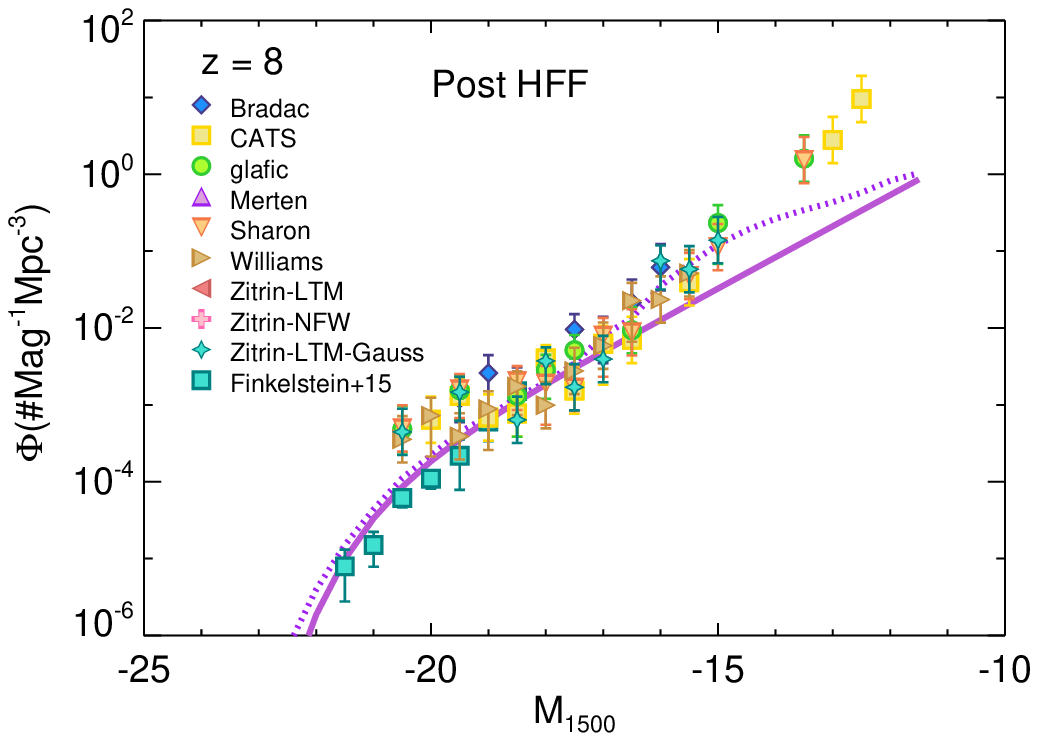}
\caption{Rest-frame UV luminosity functions based on each individual lens model, all produced in advance of the Frontier Fields data (left), or the same teams' models produced after the Frontier Fields data (right). As a guide, we also show the unlensed sample of \citet{2015ApJ...810...71F} used to fit the bright end, and the best fit Schechter function fit to our fiducial luminosity function, both intrinsic (solid) and convolved with magnitude errors (dotted). There is scatter due to the differences in magnification estimates, but all lens models show a steeply rising faint-end slope.}
\label{fig:lfprepost}
\end{figure*}

We use lens models produced both before and after the Frontier Fields data were acquired. The specific versions of each team's model used are listed in Tables \ref{tab:modelfits_z6} -- \ref{tab:modelfits_z8}, and the resulting luminosity functions for each redshift bin are shown in Figure \ref{fig:lfprepost}. There is scatter between the luminosity functions produced with different lens models as they each result in different magnification estimates for the sample galaxies. However, they all show a steeply rising faint-end slope.

We carry out the full MCMC Schechter parameter fit to each lens model individually in the same manner as for the fiducial results, and the resulting best-fitting parameters are listed in Tables \ref{tab:modelfits_z6} -- \ref{tab:modelfits_z8}.

\begin{table}
  \begin{deluxetable*}{l c c c c c c}
\centering
\tablecaption{Schechter function fit parameters from individual lens models at $z = $6 \label{tab:modelfits_z6}}
\tablecolumns{7}
\tablewidth{0pt}
\tablehead{ \colhead{Model} & \multicolumn{3}{c}{Pre-HFF} & \multicolumn{3}{c}{Post-HFF} \\
 & \colhead{$M^{\ast}$} & \colhead{$\log\phi^{\ast}$} & \colhead{$\alpha$} & \colhead{$M^{\ast}$} & \colhead{$\log\phi^{\ast}$} & \colhead{$\alpha$} }
\startdata
Bradac (v1,2$^a$/3$^b$) & $-20.830^{+0.053}_{-0.038}$ & $-3.650^{+0.041}_{-0.033}$ & $-2.10^{+0.03}_{-0.04}$ & $-20.825^{+0.046}_{-0.037}$ & $-3.644^{+0.037}_{-0.033}$ & $-2.09^{+0.03}_{-0.03}$ \\
CATS (v1,3) & $-20.822^{+0.050}_{-0.041}$ & $-3.644^{+0.038}_{-0.036}$ & $-2.10^{+0.04}_{-0.04}$ & $-20.827^{+0.049}_{-0.033}$ & $-3.648^{+0.035}_{-0.031}$ & $-2.10^{+0.03}_{-0.03}$ \\
glafic (v1$^a$,3) & $-20.827^{+0.051}_{-0.042}$ & $-3.644^{+0.037}_{-0.034}$ & $-2.09^{+0.03}_{-0.04}$ & $-20.821^{+0.048}_{-0.041}$ & $-3.645^{+0.035}_{-0.034}$ & $-2.10^{+0.03}_{-0.03}$ \\
Merten (v1) & $-20.816^{+0.043}_{-0.037}$ & $-3.638^{+0.033}_{-0.035}$ & $-2.09^{+0.03}_{-0.03}$ & - & - & - \\
Sharon (v2,3) & $-20.822^{+0.050}_{-0.040}$ & $-3.643^{+0.036}_{-0.033}$ & $-2.10^{+0.03}_{-0.03}$ & $-20.826^{+0.049}_{-0.038}$ & $-3.649^{+0.038}_{-0.033}$ & $-2.10^{+0.03}_{-0.03}$ \\
Williams (v1,3) & $-20.820^{+0.046}_{-0.037}$ & $-3.645^{+0.038}_{-0.034}$ & $-2.10^{+0.03}_{-0.03}$ & $-20.820^{+0.051}_{-0.042}$ & $-3.642^{+0.037}_{-0.037}$ & $-2.10^{+0.03}_{-0.03}$ \\
Zitrin-LTM (v1) & $-20.824^{+0.050}_{-0.039}$ & $-3.647^{+0.040}_{-0.033}$ & $-2.11^{+0.04}_{-0.03}$ & - & - & - \\
Zitrin-NFW (v1,3) & $-20.861^{+0.076}_{-0.066}$ & $-3.631^{+0.053}_{-0.047}$ & $-2.03^{+0.07}_{-0.06}$ & $-20.861^{+0.081}_{-0.079}$ & $-3.631^{+0.056}_{-0.052}$ & $-2.02^{+0.07}_{-0.07}$ \\
Zitrin-LTM-Gauss (v1$^b$,3) & $-20.829^{+0.055}_{-0.040}$ & $-3.642^{+0.041}_{-0.036}$ & $-2.08^{+0.04}_{-0.04}$ & $-20.821^{+0.047}_{-0.037}$ & $-3.644^{+0.036}_{-0.036}$ & $-2.10^{+0.03}_{-0.03}$ \\
 & & & & & & \\
\hline
Mean & $-20.828 \pm 0.013$ & $-3.643 \pm 0.005$ & $-2.09 \pm 0.02$ & $-20.829 \pm 0.014$ & $-3.643 \pm 0.006$ & $-2.09 \pm 0.03$ \\
Median & $-20.824^{+0.004}_{-0.005}$ & $-3.644^{+0.005}_{-0.003}$ & $-2.10^{+0.01}_{-0.01}$ & $-20.825^{+0.004}_{-0.003}$ & $-3.644^{+0.002}_{-0.004}$ & $-2.10^{+0.02}_{-0.01}$ \\
\enddata
\tablecomments{Best-fitting Schechter function parameters for each lens model, for the models produced by each team both before and after the Frontier Fields data. The model versions shown are those used for the (pre,post)-HFF fits.\\ $^a$ Model exists for A2744 only.\\ $^b$ Model exists for M0416 only.\\ All uncertainties are the 68\% central confidence interval, except for the error on the mean which is given as the standard deviation of all of the models.}
\end{deluxetable*}

\end{table}
\begin{table}
  \begin{deluxetable*}{l c c c c c c}
\centering
\tablecaption{Schechter function fit parameters from individual lens models at $z = $7 \label{tab:modelfits_z7}}
\tablecolumns{7}
\tablewidth{0pt}
\tablehead{ \colhead{Model} & \multicolumn{3}{c}{Pre-HFF} & \multicolumn{3}{c}{Post-HFF} \\
 & \colhead{$M^{\ast}$} & \colhead{$\log\phi^{\ast}$} & \colhead{$\alpha$} & \colhead{$M^{\ast}$} & \colhead{$\log\phi^{\ast}$} & \colhead{$\alpha$} }
\startdata
Bradac (v1,2$^a$/3$^b$) & $-20.796^{+0.050}_{-0.045}$ & $-3.673^{+0.047}_{-0.042}$ & $-2.08^{+0.04}_{-0.04}$ & $-20.799^{+0.050}_{-0.046}$ & $-3.671^{+0.043}_{-0.042}$ & $-2.08^{+0.04}_{-0.04}$ \\
CATS (v1,3) & $-20.798^{+0.053}_{-0.049}$ & $-3.671^{+0.046}_{-0.047}$ & $-2.07^{+0.04}_{-0.04}$ & $-20.815^{+0.063}_{-0.056}$ & $-3.676^{+0.052}_{-0.052}$ & $-2.05^{+0.05}_{-0.05}$ \\
glafic (v1$^a$,3) & $-20.799^{+0.053}_{-0.051}$ & $-3.669^{+0.045}_{-0.043}$ & $-2.08^{+0.05}_{-0.05}$ & $-20.806^{+0.060}_{-0.052}$ & $-3.675^{+0.049}_{-0.048}$ & $-2.05^{+0.05}_{-0.05}$ \\
Merten (v1) & $-20.805^{+0.061}_{-0.054}$ & $-3.673^{+0.051}_{-0.049}$ & $-2.05^{+0.05}_{-0.05}$ & - & - & - \\
Sharon (v2,3) & $-20.799^{+0.053}_{-0.047}$ & $-3.669^{+0.046}_{-0.044}$ & $-2.07^{+0.04}_{-0.05}$ & $-20.799^{+0.057}_{-0.049}$ & $-3.670^{+0.047}_{-0.049}$ & $-2.06^{+0.05}_{-0.04}$ \\
Williams (v1,3) & $-20.802^{+0.057}_{-0.050}$ & $-3.674^{+0.046}_{-0.046}$ & $-2.07^{+0.04}_{-0.04}$ & $-20.797^{+0.059}_{-0.047}$ & $-3.671^{+0.050}_{-0.047}$ & $-2.05^{+0.04}_{-0.04}$ \\
Zitrin-LTM (v1) & $-20.800^{+0.055}_{-0.048}$ & $-3.671^{+0.047}_{-0.046}$ & $-2.07^{+0.05}_{-0.05}$ & - & - & - \\
Zitrin-NFW (v1,3) & $-20.814^{+0.067}_{-0.062}$ & $-3.674^{+0.051}_{-0.053}$ & $-2.04^{+0.07}_{-0.07}$ & $-20.809^{+0.066}_{-0.058}$ & $-3.674^{+0.051}_{-0.058}$ & $-2.05^{+0.06}_{-0.06}$ \\
Zitrin-LTM-Gauss (v1$^b$,3) & $-20.809^{+0.066}_{-0.052}$ & $-3.676^{+0.053}_{-0.050}$ & $-2.06^{+0.06}_{-0.06}$ & $-20.806^{+0.061}_{-0.057}$ & $-3.668^{+0.051}_{-0.050}$ & $-2.05^{+0.05}_{-0.05}$ \\
 & & & & & & \\
\hline
Mean & $-20.802 \pm 0.006$ & $-3.673 \pm 0.002$ & $-2.07 \pm 0.01$ & $-20.804 \pm 0.006$ & $-3.672 \pm 0.003$ & $-2.05 \pm 0.01$ \\
Median & $-20.800^{+0.002}_{-0.008}$ & $-3.673^{+0.003}_{-0.002}$ & $-2.07^{+0.01}_{-0.01}$ & $-20.806^{+0.007}_{-0.003}$ & $-3.671^{+0.001}_{-0.004}$ & $-2.05^{+0.01}_{-0.01}$ \\
\enddata
\tablecomments{See notes to Table \ref{tab:modelfits_z6}}
\end{deluxetable*}

\end{table}
\begin{table}
  \begin{deluxetable*}{l c c c c c c}
\centering
\tablecaption{Schechter function fit parameters from individual lens models at $z = $8 \label{tab:modelfits_z8}}
\tablecolumns{7}
\tablewidth{0pt}
\tablehead{ \colhead{Model} & \multicolumn{3}{c}{Pre-HFF} & \multicolumn{3}{c}{Post-HFF} \\
 & \colhead{$M^{\ast}$} & \colhead{$\log\phi^{\ast}$} & \colhead{$\alpha$} & \colhead{$M^{\ast}$} & \colhead{$\log\phi^{\ast}$} & \colhead{$\alpha$} }
\startdata
Bradac (v1,2$^a$/3$^b$) & $-20.719^{+0.178}_{-0.140}$ & $-3.780^{+0.139}_{-0.141}$ & $-2.05^{+0.09}_{-0.09}$ & $-20.726^{+0.190}_{-0.146}$ & $-3.788^{+0.146}_{-0.152}$ & $-2.08^{+0.09}_{-0.09}$ \\
CATS (v1,3) & $-20.708^{+0.180}_{-0.146}$ & $-3.764^{+0.123}_{-0.129}$ & $-1.99^{+0.09}_{-0.09}$ & $-20.712^{+0.193}_{-0.158}$ & $-3.756^{+0.129}_{-0.133}$ & $-1.98^{+0.08}_{-0.09}$ \\
glafic (v1$^a$,3) & $-20.748^{+0.189}_{-0.135}$ & $-3.808^{+0.156}_{-0.144}$ & $-2.09^{+0.09}_{-0.10}$ & $-20.711^{+0.186}_{-0.149}$ & $-3.764^{+0.125}_{-0.131}$ & $-2.00^{+0.08}_{-0.08}$ \\
Merten (v1) & $-20.722^{+0.183}_{-0.159}$ & $-3.767^{+0.128}_{-0.132}$ & $-1.99^{+0.08}_{-0.08}$ & - & - & - \\
Sharon (v2,3) & $-20.685^{+0.199}_{-0.168}$ & $-3.745^{+0.124}_{-0.137}$ & $-1.98^{+0.09}_{-0.10}$ & $-20.743^{+0.177}_{-0.155}$ & $-3.785^{+0.135}_{-0.136}$ & $-1.99^{+0.08}_{-0.08}$ \\
Williams (v1,3) & $-20.735^{+0.180}_{-0.143}$ & $-3.790^{+0.139}_{-0.140}$ & $-2.04^{+0.08}_{-0.08}$ & $-20.718^{+0.186}_{-0.147}$ & $-3.774^{+0.136}_{-0.141}$ & $-2.02^{+0.08}_{-0.08}$ \\
Zitrin-LTM (v1) & $-20.720^{+0.189}_{-0.150}$ & $-3.780^{+0.137}_{-0.141}$ & $-2.02^{+0.09}_{-0.09}$ & - & - & - \\
Zitrin-NFW (v1,3) & $-20.745^{+0.257}_{-0.199}$ & $-3.826^{+0.190}_{-0.184}$ & $-2.02^{+0.14}_{-0.14}$ & $-20.754^{+0.260}_{-0.215}$ & $-3.830^{+0.191}_{-0.193}$ & $-2.01^{+0.15}_{-0.14}$ \\
Zitrin-LTM-Gauss (v1$^b$,3) & $-20.752^{+0.233}_{-0.180}$ & $-3.823^{+0.180}_{-0.171}$ & $-2.03^{+0.11}_{-0.10}$ & $-20.735^{+0.185}_{-0.150}$ & $-3.786^{+0.136}_{-0.139}$ & $-2.03^{+0.08}_{-0.08}$ \\
 & & & & & & \\
\hline
Mean & $-20.726 \pm 0.022$ & $-3.787 \pm 0.028$ & $-2.02 \pm 0.04$ & $-20.728 \pm 0.016$ & $-3.783 \pm 0.024$ & $-2.02 \pm 0.03$ \\
Median & $-20.722^{+0.011}_{-0.026}$ & $-3.780^{+0.015}_{-0.039}$ & $-2.02^{+0.03}_{-0.03}$ & $-20.726^{+0.014}_{-0.018}$ & $-3.785^{+0.022}_{-0.004}$ & $-2.01^{+0.02}_{-0.02}$ \\
\enddata
\tablecomments{See notes to Table \ref{tab:modelfits_z6}}
\end{deluxetable*}

\end{table}

The different lens models do not have a strong effect on the fits of $M^{\ast}$ and $\phi^{\ast}$ at $z = 6$ or 7 as the bright end is constrained by the unlensed CANDELS data. There are differences between the models at $z \sim 8$, where the overdensity of bright galaxies in Abell 2744 affects the fit.

We expect the lens model uncertainties to affect the measurement of $\alpha$, which is primarily constrained by lensed galaxies, and we do find a wider range of values of $\alpha$ by looking at the different lens models. However, every lens model - both those produced before and after the HFF data - is consistent with the fiducial result to within $1\,\sigma$. We find no systematic difference between the models that use the light-traces-mass assumption and those that do not. At $z \sim 6$, the differences between the pre- and post-HFF models are statistically insignificant, and all of the models are in agreement within 1$\sigma$. At $z \sim 7$ the post-HFF versions of the models produce systematically shallower faint-end slopes than the originals, but the differences here and at $z \sim 8$ are again all within $1\,\sigma$ of the original models.

We conclude that although the different lens models indicate large differences in the magnifications of individual galaxies, the effect on the overall luminosity function is minimal. We caution, however, that although the different lens models cover different assumptions and methodologies, they all use the same input data, so this comparison is not indicative of the full errors in the lens modelling. However, it does provide an estimate of the uncertainties resulting from the lens modelling methodology, and we incorporate this as a systematic uncertainty in our fiducial Schechter function fits in Table \ref{tab:lfpars}.

\section{Discussion}
\label{sec:disc}

The discovery of these extremely faint galaxies allows us to place much tighter constraints on $\alpha$ than is possible with unlensed surveys, to a fractional error $\sigma_{\alpha}$/$\alpha  <$2\% at $z =$ 6 and 7.  Crucially, at $z =$ 6, we extend direct observations of galaxies to luminosities more than a factor of 100 times deeper than the unlensed \textit{HST} limits \citep{2015ApJ...810...71F,2015ApJ...803...34B}, and more than a factor of 10 times deeper (to $M_\mathrm{UV} = -12.5$) than previous work with these data \citep{2015ApJ...814...69AA}.  The faintest galaxies in our sample are the progenitors of today's galaxies similar to the Fornax dwarf spheroidal \citep{2015MNRAS.453.1503B}.

Interestingly, our observations show a faint-end slope that is consistent with a single power-law to these extreme faint luminosities, with no apparent turnover.  This is critical, as previous models of reionization have found that the luminosity function must continue unbroken to at least M$_{UV} = -$13 to complete reionization by $z \sim 6$ \citep{2015ApJ...810...71F,2015ApJ...802L..19R}.  Our observations now provide the first empirical evidence that this is the case.

Given the increased luminosity distance to higher redshifts, our faintest galaxies are somewhat intrinsically brighter,  to $M_\mathrm{UV} = -14.5$ at $z \sim$ 7, and $M_\mathrm{UV} = -15$ at $z \sim$ 8.  These are a factor of $3\times$ and $17\times$ fainter in luminosity than previous studies at $z \sim$ 7 and 8, respectively.  Although fainter galaxies at $z \sim$ 7 and 8 than we observe are needed to complete reionization by these earlier times, the presence of such galaxies at $z \sim$ 6 strongly implies that future deep lensing surveys with the {\it James Webb Space Telescope} will find them at $z \geq$ 7.

By maximizing the power of gravitational lensing, our observed sample of faint galaxies has now approached the magnitude range where simulations predict a turnover in the luminosity function.  Had such a turnover been found, it would reduce the previously extrapolated luminosity density, and thus would cast doubt on the ability of star-forming galaxies to reionize the universe.

In order to consider whether our data rule out a turnover at the faint end of the luminosity function, we adopt the modified Schechter function of \citet{2013ApJ...766...94J}

\begin{equation}
\phi\left(L\right) = \phi^{\ast}\left(\frac{L}{L^{\ast}}\right)^{\alpha}\exp\left(-\frac{L}{L^{\ast}}\right)\left[1 + \left(\frac{L}{L_t}\right)^{\beta}\right]^{-1}
\label{eq:modschechter}
\end{equation}

where $L^{\ast}$, $\phi^{\ast}$ and $\alpha$ have their usual meanings, $L_t$ is the luminosity of the turnover and $\beta$ represents the slope after the turnover, such that $\phi\left(L\right) \propto L^{\alpha - \beta}$ when $L \ll L_t$. For simplicity, we consider a flattened faint end such that $\beta = \alpha$.

Using the magnitude form of Equation \ref{eq:modschechter}, we keep the best-fit parameters $M^{\ast}$, $\phi^{\ast}$ and $\alpha$ from our fiducial luminosity function fit and calculate the likelihood of the turnover magnitude $M_t$ at a range of magnitudes $-16 \leq M_t \leq 10$. To compare the results, we use the Bayesian Information Criterion \citep[BIC;][]{2004MNRAS.351L..49L}, defined as

\begin{equation}
\rm{BIC} = -2\ln\left(\mathcal{L}\right) + k\ln\left(N\right)
\label{eq:bic}
\end{equation}

where $\mathcal{L}$ is the likelihood of the observed $\phi\left(M\right)$ given the modified Schechter function parameters, $k$ is the number of free parameters and $N$ is the number of data points. A lower BIC indicates a preferred model. The difference in BIC, $\Delta$BIC, between the modified Schechter function and that of the fiducial fit with no turnover is calculated for each value of $M_t$.

A $\Delta$BIC $> 10$ indicates very strong evidence against the model with the higher BIC. A value $6 < \Delta\rm{BIC} < 10$ indicates strong evidence against the model with higher BIC, and $2 < \Delta\rm{BIC} < 6$ indicates positive evidence against the higher BIC. If $\Delta\rm{BIC} < 2$, there is no significant evidence in favor of either model.

\begin{figure}
\includegraphics[width=0.5\textwidth]{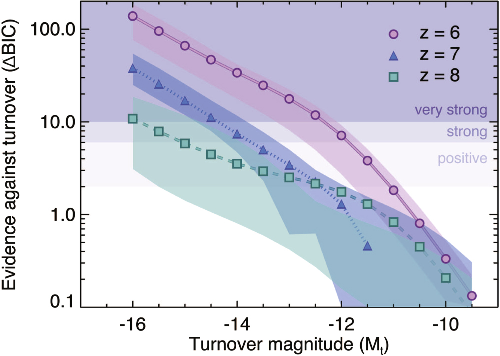}
\caption{The difference in Bayes Information Criterion (BIC) between the fiducial Schechter function fit and a modified Schechter function incorporating a turnover at $M_t$. In all cases, the modified Schechter function gives a higher BIC (worse fit) than the fiducial model. The points are the results when using the median magnification for each galaxy in the sample, and the shaded regions show the full range when each available lens model is used individually. The highlighted horizontal regions show the values of $\Delta$BIC that indicate positive, strong or very strong evidence against the modified model.}
\label{fig:turnover}
\end{figure}

We show the $\Delta$BIC results for the full range of $M_t$ in Figure \ref{fig:turnover}. We find the BIC is higher for all of the modified Schechter functions than for the fiducial model at all $M_t$, but $\Delta$BIC approaches zero at the faint end where we have no data.

The $\Delta$BIC indicates very strong evidence against a turnover at $M_t < -12.5$ at $z \sim 6$, $M_t < -14.5$ at $z\sim 7$ and $M_t < -16$ at $z \sim 8$. Strong evidence is indicated for $M_t < -12$, $M_t < -13.5$ and $M_t < -15.5$, and there is positive evidence against $M_t < -11$, $M_t < -12$ and $M_t < -13$ at $z \sim$ 6, 7 and 8 respectively. While the $z \sim 6$ results may somewhat be driven by the very faint intrinsic luminosity of our faintest source, the model with the lowest magnification (the Zitrin-LTM-Gauss version 3 model, with $\mu = 22.6$) would place this galaxy at $M_{\rm{UV}} = -14.1$. To explore the effect of this uncertainty, we calculate the $\Delta$BIC between our fiducial Schechter function and the modified function with a turnover omitting this galaxy entirely, and find that there is still very strong evidence against a turnover at magnitudes brighter than $-12.5$.

These results are in modest tension with a number of recent theoretical results \citep{2015ApJ...807L..12O,2013ApJ...766...94J,2015arXiv151200563L} and with constraints from local dwarf galaxies \citep{2014MNRAS.443L..44B,2015MNRAS.453.1503B}, implying that star formation is proceeding in lower-mass halos than these simulations suggest. However, some other work suggests that stars could still form in these low-mass halos \citep{2011MNRAS.410.1703F,2016arXiv160401314Y,2016arXiv160307729G}. If the luminosity function does continue to rise to $M_{\rm{UV}} = -13$, this does support the integration of the luminosity function to this limit for the purposes of calculating the ionizing radiation available for reionization. Therefore, these results support the idea that faint galaxies in the early Universe could have been responsible for reionization.\\
\\

\section{Conclusions}
\label{sec:conc}

We have developed a new technique using wavelet decomposition to subtract intracluster light and foreground galaxies from deep \emph{HST} imaging, enabling the detection of faint, highly magnified galaxies in the first two Hubble Frontier Fields clusters.

Having tested that this method enables the recovery of faint galaxies with accurate photometry, we used it to select 167 galaxies at $z \gtrsim 6$ in the Abell 2744 and MACS 0416 fields.

We carried out detailed completeness simulations to calculate the effective volume of our survey and used this to compute the UV luminosity functions. The high magnifications of some of these sources mean that we are able to constrain the luminosity function to fainter limits than previous work; to $M_\mathrm{UV} = -12.5$ at $z \sim$ 6, $M_\mathrm{UV} = -14.5$ at $z \sim$ 7, and $M_\mathrm{UV} = -15$ at $z \sim$ 8, a factor of $12\times$, $3\times$ and $17\times$ fainter in luminosity than previous studies at $z =$ 6, 7 and 8, respectively. We have found steep faint-end slopes ($\alpha < -2$) extending at least to the limit of our observations at each redshift, with no evidence of a turnover. This is critical, as previous models of reionization have found that the luminosity function must continue unbroken to at least M$_{UV} = -$13 to complete reionization by a redshift of six \citep{2015ApJ...810...71F,2015ApJ...802L..19R}.  Our observations now provide the first empirical evidence that this is the case at $z \sim 6$.

We also investigated the systematic differences between the available lens models. We find that although they imply large differences in the magnification factors of individual galaxies, the effect on the luminosity function is small, with fractional uncertainties on the faint-end slope $\alpha$ of $\sigma_{\alpha}/\alpha < 4\%$. We find no systematic difference in the best-fitting $\alpha$ between the lens models that use the light-traces-mass assumption and those that do not. After being updated with additional constraints from the Frontier Fields data, the lens models all produce results within $1\,\sigma$ of the original models.

We have also carried out a statistical Bayesian analysis to determine which turnover magnitudes can be ruled out our results. We have found very strong evidence against a turnover at magnitudes brighter than $M_\mathrm{UV} = -$12.5 at $z =$ 6, and strong evidence against a turnover at magnitudes brighter than $M_\mathrm{UV} = -$13.5 and $-$15 at $z =$ 7 and 8, respectively.

Our study demonstrates the power of gravitational lensing to study the early Universe and directly observe the very faintest galaxies that powered reionization. The complete Hubble Frontier Fields program will further constrain the properties of these early galaxies, providing a preview of the science that will be possible with the \textit{James Webb Space Telescope} and future ground and space-based facilities.

\acknowledgments

The authors would like to thank the anonymous referee for suggestions that improved this paper. We also thank Anton Koekemoer and his team for their work in reducing and making available the Hubble Frontier Fields data, Hakim Atek for providing useful information on his catalogs and Mathilde Jauzac for providing updated lens models in advance of their publication. This work is based on observations obtained with the NASA/ESA Hubble Space Telescope, retrieved from the Mikulski Archive for Space Telescopes (MAST) at the Space Telescope Science Institute (STScI). This work utilizes gravitational lensing models produced by PIs Brada\u{c}, Ebeling, Merten \& Zitrin, Sharon, and Williams funded as part of the HST Frontier Fields program conducted by STScI. STScI is operated by the Association of Universities for Research in Astronomy, Inc. under NASA contract NAS 5-26555. The lens models were obtained from the Mikulski Archive for Space Telescopes (MAST). The authors acknowledge the Texas Advanced Computing Center (TACC) at The University of Texas at Austin for providing HPC resources that have contributed to the research results reported within this paper. URL: http://www.tacc.utexas.edu. R.C.L. acknowledges support from the University of Texas at Austin and NASA through grant number AR-14300 from the Space Telescope Science Institute, which is operated by AURA, Inc. under NASA contract NAS 5-26555. S.L.F.  acknowledges support from the University of Texas at Austin.

%% To help institutions obtain information on the effectiveness of their
%% telescopes, the AAS Journals has created a group of keywords for telescope
%% facilities. A common set of keywords will make these types of searches
%% significantly easier and more accurate. In addition, they will also be
%% useful in linking papers together which utilize the same telescopes
%% within the framework of the National Virtual Observatory.
%% See the AASTeX Web site at http://aastex.aas.org/
%% for information on obtaining the facility keywords.

%% After the acknowledgments section, use the following syntax and the
%% \facility{} macro to list the keywords of facilities used in the research
%% for the paper.  Each keyword will be checked against the master list during
%% copy editing.  Individual instruments or configurations can be provided 
%% in parentheses, after the keyword, but they will not be verified.

{\it Facilities:} \facility{HST (ACS, WFC3)}.

\bibliographystyle{aasjournal}
\bibliography{bib}

\vspace*{6in}

\appendix

\section{Catalogs}

\begin{deluxetable*}{l r r c c c}
\tablewidth{0pt}
\tablecaption{Catalog of the z$\sim$6 sample \label{tab:z6}}
\tablehead{\colhead{ID} & \colhead{RA} & \colhead{Dec} & \colhead{$z_{\mathrm{phot}}^a$} & \colhead{Magnification$^b$} & \colhead{$M_{1500}^c$} \\
 & \colhead{(J2000)} & \colhead{(J2000)} & & & \colhead{(intrinsic AB mag)}}
\startdata
A2744\_z6\_2968 & 3.589725 & -30.403378 & 5.34$^{+1.18}_{-0.57}$ & 2.7$^{+0.8}_{-1.1}$ & $-16.37 ^{+0.40}_{-0.32}$ \\
A2744\_z6\_2101 & 3.591432 & -30.396690 & 5.59$^{+0.08}_{-0.28}$ & 13.2$^{+4.3}_{-6.8}$ & $-16.80 ^{+0.45}_{-0.31}$ \\
A2744\_z6\_197 & 3.590764 & -30.379410 & 5.59$^{+0.09}_{-0.22}$ & 1.7$^{+1.2}_{-0.6}$ & $-19.30 ^{+0.35}_{-0.60}$ \\
A2744\_z6\_1506 & 3.590724 & -30.391712 & 5.61$^{+1.83}_{-0.48}$ & 11.1$^{+6.5}_{-2.7}$ & $-14.74 ^{+0.34}_{-0.54}$ \\
A2744\_z6\_203431 & 3.595881 & -30.400272 & 5.61$^{+0.50}_{-0.88}$ & 4.7$^{+2.1}_{-2.7}$ & $-16.80 ^{+0.52}_{-0.44}$ \\
A2744\_z6\_21947 & 3.583411 & -30.400023 & 5.63$^{+0.40}_{-0.41}$ & 15.8$^{+14.3}_{-6.3}$ & $-15.39 ^{+0.39}_{-0.71}$ \\
M0416\_z6\_1672 & 64.046310 & -24.069864 & 5.65$^{+0.17}_{-0.44}$ & 4.4$^{+1.3}_{-1.3}$ & $-17.45 ^{+0.30}_{-0.30}$ \\
A2744\_z6\_11116 & 3.585763 & -30.390440 & 5.67$^{+0.43}_{-1.30}$ & 9.7$^{+3.1}_{-3.8}$ & $-15.31 ^{+0.40}_{-0.36}$ \\
A2744\_z6\_184008 & 3.597625 & -30.400465 & 5.67$^{+0.68}_{-0.93}$ & 4.2$^{+1.5}_{-2.2}$ & $-16.53 ^{+0.49}_{-0.39}$ \\
A2744\_z6\_596 & 3.603425 & -30.383221 & 5.67$^{+0.12}_{-0.30}$ & 1.4$^{+0.6}_{-0.5}$ & $-18.47 ^{+0.34}_{-0.39}$ \\
M0416\_z6\_4754 & 64.034515 & -24.094074 & 5.70$^{+0.14}_{-0.66}$ & 1.3$^{+0.4}_{-0.2}$ & $-18.29 ^{+0.20}_{-0.30}$ \\
A2744\_z6\_23196 & 3.591314 & -30.412609 & 5.71$^{+0.44}_{-0.62}$ & 5.4$^{+2.3}_{-1.7}$ & $-16.02 ^{+0.33}_{-0.42}$ \\
M0416\_z6\_4827 & 64.042404 & -24.095964 & 5.71$^{+0.22}_{-0.99}$ & 1.3$^{+0.1}_{-0.2}$ & $-17.11 ^{+0.21}_{-0.20}$ \\
M0416\_z6\_1410 & 64.041046 & -24.067846 & 5.71$^{+0.14}_{-0.92}$ & 9.3$^{+5.0}_{-4.1}$ & $-16.04 ^{+0.42}_{-0.48}$ \\
A2744\_z6\_3321 & 3.597102 & -30.405430 & 5.72$^{+0.23}_{-1.09}$ & 15.4$^{+5.2}_{-6.4}$ & $-15.35 ^{+0.39}_{-0.34}$ \\
A2744\_z6\_33756 & 3.585379 & -30.411974 & 5.72$^{+0.35}_{-0.85}$ & 3.5$^{+1.8}_{-1.1}$ & $-17.46 ^{+0.33}_{-0.47}$ \\
M0416\_z6\_183 & 64.042938 & -24.057184 & 5.72$^{+0.11}_{-0.23}$ & 2.8$^{+0.3}_{-1.3}$ & $-17.28 ^{+0.41}_{-0.12}$ \\
A2744\_z6\_3567 & 3.605219 & -30.407095 & 5.72$^{+0.25}_{-0.89}$ & 2.4$^{+1.4}_{-0.9}$ & $-16.97 ^{+0.38}_{-0.51}$ \\
M0416\_z6\_30052 & 64.026199 & -24.054296 & 5.73$^{+0.19}_{-0.90}$ & 1.3$^{+0.1}_{-0.3}$ & $-17.15 ^{+0.28}_{-0.22}$ \\
M0416\_z6\_107 & 64.032150 & -24.055706 & 5.75$^{+0.10}_{-1.02}$ & 1.7$^{+0.2}_{-0.6}$ & $-17.53 ^{+0.35}_{-0.16}$ \\
M0416\_z6\_103 & 64.032455 & -24.055666 & 5.75$^{+0.12}_{-0.37}$ & 1.7$^{+0.2}_{-0.6}$ & $-17.92 ^{+0.35}_{-0.15}$ \\
A2744\_z6\_22178 & 3.597490 & -30.402424 & 5.76$^{+0.48}_{-0.74}$ & 6.0$^{+1.7}_{-3.2}$ & $-15.46 ^{+0.48}_{-0.32}$ \\
A2744\_z6\_3860 & 3.604751 & -30.409290 & 5.77$^{+0.10}_{-0.93}$ & 2.7$^{+1.8}_{-1.0}$ & $-17.85 ^{+0.36}_{-0.55}$ \\
A2744\_z6\_4290 & 3.592946 & -30.413328 & 5.78$^{+0.09}_{-0.60}$ & 8.4$^{+2.7}_{-2.1}$ & $-16.32 ^{+0.25}_{-0.30}$ \\
A2744\_z6\_1848 & 3.593192 & -30.394682 & 5.78$^{+1.08}_{-1.15}$ & 7.2$^{+2.0}_{-1.2}$ & $-15.25 ^{+0.34}_{-0.39}$ \\
M0416\_z6\_112879 & 64.039886 & -24.073895 & 5.80$^{+0.25}_{-0.77}$ & 19.0$^{+6.7}_{-13.0}$ & $-14.23 ^{+0.58}_{-0.35}$ \\
A2744\_z6\_115310 & 3.569330 & -30.407413 & 5.80$^{+0.17}_{-0.98}$ & 1.6$^{+0.7}_{-0.6}$ & $-17.30 ^{+0.38}_{-0.39}$ \\
M0416\_z6\_706 & 64.042244 & -24.063688 & 5.80$^{+0.09}_{-0.15}$ & 2.4$^{+1.3}_{-0.8}$ & $-18.76 ^{+0.32}_{-0.47}$ \\
M0416\_z6\_2037 & 64.026894 & -24.072798 & 5.80$^{+0.15}_{-1.03}$ & 5.0$^{+0.5}_{-1.9}$ & $-16.50 ^{+0.36}_{-0.16}$ \\
A2744\_z6\_3032 & 3.570067 & -30.403719 & 5.80$^{+0.10}_{-0.35}$ & 1.7$^{+1.1}_{-0.7}$ & $-18.92 ^{+0.37}_{-0.54}$ \\
A2744\_z6\_116221 & 3.582199 & -30.413094 & 5.81$^{+0.47}_{-1.13}$ & 2.8$^{+0.6}_{-1.2}$ & $-16.32 ^{+0.42}_{-0.27}$ \\
M0416\_z6\_603 & 64.054520 & -24.062000 & 5.81$^{+0.11}_{-0.79}$ & 2.2$^{+1.3}_{-1.0}$ & $-16.96 ^{+0.43}_{-0.52}$ \\
A2744\_z6\_3351 & 3.573700 & -30.405655 & 5.82$^{+0.35}_{-1.33}$ & 2.1$^{+0.3}_{-0.8}$ & $-16.67 ^{+0.38}_{-0.23}$ \\
A2744\_z6\_3843 & 3.586703 & -30.409325 & 5.82$^{+0.45}_{-1.00}$ & 6.8$^{+1.7}_{-3.5}$ & $-15.37 ^{+0.48}_{-0.30}$ \\
M0416\_z6\_1275 & 64.042580 & -24.066776 & 5.84$^{+0.43}_{-1.30}$ & 5.1$^{+14.3}_{-1.0}$ & $-15.69 ^{+0.28}_{-1.45}$ \\
A2744\_z6\_33454 & 3.581096 & -30.408943 & 5.84$^{+0.40}_{-0.55}$ & 3.7$^{+0.9}_{-2.5}$ & $-16.27 ^{+0.57}_{-0.29}$ \\
A2744\_z6\_121854 & 3.586404 & -30.395344 & 5.84$^{+0.91}_{-0.81}$ & 3.7$^{+1.1}_{-2.9}$ & $-16.32 ^{+0.65}_{-0.34}$ \\
A2744\_z6\_227 & 3.590541 & -30.379765 & 5.85$^{+0.18}_{-0.47}$ & 1.7$^{+1.3}_{-0.6}$ & $-18.06 ^{+0.35}_{-0.60}$ \\
M0416\_z6\_4250 & 64.039551 & -24.088539 & 5.86$^{+0.04}_{-0.33}$ & 1.7$^{+0.2}_{-0.4}$ & $-19.57 ^{+0.26}_{-0.13}$ \\
A2744\_z6\_23168 & 3.591140 & -30.412184 & 5.89$^{+0.74}_{-0.71}$ & 8.3$^{+2.4}_{-2.3}$ & $-15.33 ^{+0.37}_{-0.38}$ \\
M0416\_z6\_1209 & 64.050858 & -24.066536 & 5.89$^{+0.12}_{-0.17}$ & 3.4$^{+0.6}_{-1.7}$ & $-17.28 ^{+0.44}_{-0.18}$ \\
M0416\_z6\_482 & 64.047134 & -24.061138 & 5.89$^{+0.09}_{-0.25}$ & 14.4$^{+26.0}_{-7.2}$ & $-15.84 ^{+0.44}_{-1.12}$ \\
M0416\_z6\_164 & 64.043915 & -24.056280 & 5.89$^{+0.14}_{-1.25}$ & 2.6$^{+0.4}_{-1.0}$ & $-16.48 ^{+0.40}_{-0.27}$ \\
M0416\_z6\_1509 & 64.054161 & -24.068598 & 5.91$^{+0.20}_{-1.26}$ & 1.9$^{+0.2}_{-0.6}$ & $-17.28 ^{+0.35}_{-0.23}$ \\
M0416\_z6\_3661 & 64.049240 & -24.084097 & 5.91$^{+0.54}_{-1.28}$ & 1.4$^{+0.2}_{-0.3}$ & $-17.08 ^{+0.31}_{-0.27}$ \\
A2744\_z6\_989 & 3.606227 & -30.386646 & 5.92$^{+0.07}_{-0.44}$ & 1.9$^{+0.5}_{-0.6}$ & $-19.34 ^{+0.28}_{-0.27}$ \\
A2744\_z6\_2363 & 3.579679 & -30.398680 & 5.92$^{+1.26}_{-0.91}$ & 11.3$^{+13.3}_{-6.4}$ & $-14.94 ^{+0.53}_{-0.86}$ \\
M0416\_z6\_3912 & 64.015404 & -24.085691 & 5.93$^{+0.28}_{-1.32}$ & 2.3$^{+0.7}_{-0.7}$ & $-16.83 ^{+0.32}_{-0.33}$ \\
M0416\_z6\_1217 & 64.051086 & -24.066515 & 5.94$^{+0.14}_{-0.58}$ & 3.4$^{+0.5}_{-1.7}$ & $-16.93 ^{+0.44}_{-0.18}$ \\
M0416\_z6\_243 & 64.037361 & -24.057997 & 5.94$^{+0.16}_{-0.22}$ & 2.4$^{+0.3}_{-1.1}$ & $-17.17 ^{+0.42}_{-0.17}$ \\
M0416\_z6\_648 & 64.048180 & -24.062405 & 5.94$^{+0.12}_{-0.17}$ & 11.1$^{+94.8}_{-8.4}$ & $-16.04 ^{+0.61}_{-2.44}$ \\
M0416\_z6\_21350 & 64.029488 & -24.071123 & 5.95$^{+0.43}_{-0.81}$ & 9.1$^{+10.2}_{-4.6}$ & $-15.55 ^{+0.47}_{-0.82}$ \\
M0416\_z6\_518 & 64.037483 & -24.061235 & 5.95$^{+0.08}_{-0.20}$ & 3.9$^{+1.0}_{-1.2}$ & $-17.39 ^{+0.30}_{-0.25}$ \\
A2744\_z6\_82888 & 3.579229 & -30.404310 & 5.97$^{+1.44}_{-0.94}$ & 4.3$^{+1.0}_{-1.0}$ & $-16.47 ^{+0.23}_{-0.23}$ \\
M0416\_z6\_10312 & 64.030647 & -24.059616 & 5.97$^{+0.53}_{-1.38}$ & 1.8$^{+0.1}_{-0.7}$ & $-16.86 ^{+0.41}_{-0.26}$ \\
M0416\_z6\_333 & 64.044472 & -24.059307 & 5.97$^{+0.14}_{-0.25}$ & 7.5$^{+1.0}_{-5.1}$ & $-16.68 ^{+0.57}_{-0.16}$ \\
M0416\_z6\_4109 & 64.035332 & -24.087257 & 5.98$^{+0.19}_{-0.69}$ & 1.9$^{+0.6}_{-0.6}$ & $-17.80 ^{+0.30}_{-0.30}$ \\
M0416\_z6\_2517 & 64.050583 & -24.076141 & 5.99$^{+0.25}_{-1.16}$ & 1.7$^{+0.2}_{-0.4}$ & $-16.95 ^{+0.29}_{-0.23}$ \\
M0416\_z6\_4273 & 64.038712 & -24.088648 & 5.99$^{+0.51}_{-1.04}$ & 1.7$^{+0.2}_{-0.5}$ & $-17.12 ^{+0.32}_{-0.24}$ \\
A2744\_z6\_1062 & 3.570582 & -30.385790 & 6.00$^{+0.25}_{-0.34}$ & 3.3$^{+0.7}_{-1.4}$ & $-16.96 ^{+0.39}_{-0.22}$ \\
A2744\_z6\_3867 & 3.604565 & -30.409365 & 6.00$^{+0.21}_{-0.59}$ & 2.8$^{+1.8}_{-1.0}$ & $-16.50 ^{+0.35}_{-0.53}$ \\
M0416\_z6\_1823 & 64.049294 & -24.070936 & 6.01$^{+0.18}_{-0.53}$ & 2.2$^{+0.3}_{-1.0}$ & $-17.40 ^{+0.41}_{-0.18}$ \\
M0416\_z6\_4405 & 64.034554 & -24.089708 & 6.01$^{+0.08}_{-0.58}$ & 2.0$^{+0.4}_{-0.5}$ & $-17.84 ^{+0.27}_{-0.23}$ \\
A2744\_z6\_2073 & 3.586095 & -30.396465 & 6.01$^{+0.39}_{-1.37}$ & 5.1$^{+11.4}_{-1.8}$ & $-16.05 ^{+0.35}_{-1.27}$ \\
M0416\_z6\_4074 & 64.038498 & -24.087042 & 6.02$^{+0.18}_{-0.82}$ & 1.7$^{+0.3}_{-0.5}$ & $-17.84 ^{+0.31}_{-0.20}$ \\
M0416\_z6\_528 & 64.047859 & -24.062088 & 6.05$^{+0.06}_{-0.19}$ & 16.7$^{+17.0}_{-13.2}$ & $-16.68 ^{+0.63}_{-0.76}$ \\
A2744\_z6\_144225 & 3.593991 & -30.410473 & 6.05$^{+0.52}_{-0.36}$ & 14.9$^{+4.9}_{-7.6}$ & $-15.76 ^{+0.46}_{-0.34}$ \\
M0416\_z6\_10233 & 64.033836 & -24.058512 & 6.05$^{+0.51}_{-0.95}$ & 2.0$^{+0.2}_{-0.7}$ & $-16.88 ^{+0.40}_{-0.26}$ \\
M0416\_z6\_122330 & 64.050888 & -24.074293 & 6.06$^{+0.38}_{-0.80}$ & 2.0$^{+0.2}_{-0.5}$ & $-16.74 ^{+0.31}_{-0.22}$ \\
A2744\_z6\_3928 & 3.603929 & -30.409842 & 6.07$^{+0.85}_{-1.38}$ & 4.5$^{+1.8}_{-0.8}$ & $-16.02 ^{+0.32}_{-0.43}$ \\
M0416\_z6\_4851 & 64.046959 & -24.097439 & 6.09$^{+0.29}_{-1.05}$ & 1.2$^{+0.1}_{-0.1}$ & $-17.86 ^{+0.22}_{-0.20}$ \\
M0416\_z6\_140523 & 64.040039 & -24.061846 & 6.10$^{+0.13}_{-0.16}$ & 11.5$^{+6.3}_{-6.6}$ & $-16.35 ^{+0.49}_{-0.48}$ \\
A2744\_z6\_2830 & 3.584936 & -30.402237 & 6.11$^{+1.05}_{-1.22}$ & 110.0$^{+129.0}_{-22.2}$ & $-12.42 ^{+0.32}_{-0.86}$ \\
A2744\_z6\_2663 & 3.587437 & -30.401377 & 6.11$^{+0.09}_{-0.23}$ & 3.0$^{+0.4}_{-0.9}$ & $-18.34 ^{+0.28}_{-0.15}$ \\
M0416\_z6\_10645 & 64.049232 & -24.063351 & 6.11$^{+0.15}_{-0.20}$ & 5.2$^{+207.8}_{-1.9}$ & $-16.33 ^{+0.35}_{-4.03}$ \\
A2744\_z6\_143856 & 3.594747 & -30.407383 & 6.12$^{+0.56}_{-1.18}$ & 2.7$^{+1.0}_{-0.9}$ & $-17.08 ^{+0.36}_{-0.38}$ \\
A2744\_z6\_830 & 3.608999 & -30.385279 & 6.13$^{+0.22}_{-0.28}$ & 1.4$^{+0.9}_{-0.5}$ & $-17.96 ^{+0.33}_{-0.55}$ \\
M0416\_z6\_3639 & 64.030815 & -24.083683 & 6.15$^{+0.17}_{-0.31}$ & 13.0$^{+4.1}_{-4.0}$ & $-16.03 ^{+0.30}_{-0.31}$ \\
A2744\_z6\_3341 & 3.585820 & -30.405579 & 6.16$^{+0.26}_{-1.64}$ & 24.6$^{+30.6}_{-6.7}$ & $-14.33 ^{+0.30}_{-0.88}$ \\
M0416\_z6\_51068 & 64.043427 & -24.058943 & 6.19$^{+0.25}_{-0.33}$ & 4.8$^{+0.4}_{-2.8}$ & $-16.47 ^{+0.52}_{-0.16}$ \\
A2744\_z6\_10850 & 3.595286 & -30.387192 & 6.21$^{+0.92}_{-0.87}$ & 2.4$^{+2.0}_{-0.7}$ & $-16.98 ^{+0.38}_{-0.67}$ \\
A2744\_z6\_184783 & 3.601101 & -30.403963 & 6.21$^{+0.18}_{-0.18}$ & 3.4$^{+6.1}_{-0.9}$ & $-17.18 ^{+0.25}_{-1.11}$ \\
A2744\_z6\_1391 & 3.596868 & -30.390459 & 6.22$^{+0.55}_{-1.27}$ & 4.2$^{+2.3}_{-0.7}$ & $-16.41 ^{+0.26}_{-0.51}$ \\
A2744\_z6\_33246 & 3.582345 & -30.407299 & 6.22$^{+0.33}_{-0.68}$ & 4.5$^{+1.1}_{-2.5}$ & $-16.34 ^{+0.50}_{-0.28}$ \\
A2744\_z6\_45268 & 3.603210 & -30.410357 & 6.23$^{+0.16}_{-0.17}$ & 4.0$^{+0.9}_{-1.2}$ & $-17.86 ^{+0.29}_{-0.22}$ \\
A2744\_z6\_94569 & 3.599376 & -30.414194 & 6.27$^{+0.45}_{-0.70}$ & 7.0$^{+7.4}_{-2.5}$ & $-16.26 ^{+0.37}_{-0.80}$ \\
M0416\_z6\_114509 & 64.049919 & -24.082909 & 6.27$^{+0.54}_{-1.16}$ & 1.4$^{+0.2}_{-0.3}$ & $-17.17 ^{+0.29}_{-0.25}$ \\
A2744\_z6\_32567 & 3.582077 & -30.402222 & 6.31$^{+0.38}_{-0.88}$ & 11.6$^{+3.5}_{-3.4}$ & $-15.09 ^{+0.33}_{-0.28}$ \\
A2744\_z6\_3979 & 3.600618 & -30.410294 & 6.31$^{+0.17}_{-0.18}$ & 11.4$^{+2.2}_{-3.4}$ & $-15.99 ^{+0.33}_{-0.33}$ \\
M0416\_z6\_10942 & 64.050682 & -24.065805 & 6.32$^{+0.80}_{-0.84}$ & 5.4$^{+1.1}_{-2.9}$ & $-15.73 ^{+0.29}_{-0.20}$ \\
M0416\_z6\_41245 & 64.038887 & -24.060623 & 6.32$^{+0.21}_{-0.21}$ & 4.7$^{+1.2}_{-2.0}$ & $-17.04 ^{+0.50}_{-0.30}$ \\
A2744\_z6\_3762 & 3.575479 & -30.408585 & 6.34$^{+0.45}_{-0.87}$ & 2.2$^{+0.4}_{-1.1}$ & $-16.78 ^{+0.39}_{-0.26}$ \\
A2744\_z6\_454 & 3.580720 & -30.381990 & 6.36$^{+0.26}_{-0.42}$ & 2.6$^{+0.4}_{-1.0}$ & $-17.49 ^{+0.47}_{-0.23}$ \\
M0416\_z6\_4863 & 64.045547 & -24.098230 & 6.37$^{+0.35}_{-0.61}$ & 1.2$^{+0.1}_{-0.2}$ & $-17.60 ^{+0.35}_{-0.19}$ \\
M0416\_z6\_856 & 64.032951 & -24.064079 & 6.37$^{+0.86}_{-1.48}$ & 3.2$^{+0.2}_{-1.3}$ & $-16.54 ^{+0.22}_{-0.19}$ \\
A2744\_z6\_181733 & 3.600831 & -30.388754 & 6.40$^{+0.63}_{-0.99}$ & 2.2$^{+0.4}_{-0.8}$ & $-17.85 ^{+0.42}_{-0.24}$ \\
A2744\_z6\_434 & 3.605586 & -30.381670 & 6.40$^{+0.63}_{-1.57}$ & 1.3$^{+0.6}_{-0.5}$ & $-17.38 ^{+0.41}_{-0.30}$ \\
A2744\_z6\_11421 & 3.594052 & -30.393772 & 6.42$^{+0.37}_{-0.55}$ & 3.7$^{+4.1}_{-0.9}$ & $-16.98 ^{+0.40}_{-0.46}$ \\
A2744\_z6\_3623 & 3.604972 & -30.407511 & 6.42$^{+1.05}_{-1.43}$ & 3.8$^{+1.5}_{-1.0}$ & $-16.20 ^{+0.28}_{-0.82}$ \\
M0416\_z6\_3484 & 64.047829 & -24.082790 & 6.44$^{+0.20}_{-0.25}$ & 1.4$^{+0.2}_{-0.3}$ & $-18.77 ^{+0.37}_{-0.44}$ \\
M0416\_z6\_4522 & 64.025352 & -24.090925 & 6.45$^{+1.06}_{-1.55}$ & 3.3$^{+0.9}_{-1.8}$ & $-16.37 ^{+0.75}_{-0.76}$ \\
M0416\_z6\_1997 & 64.039909 & -24.072279 & 6.46$^{+0.38}_{-0.61}$ & 19.7$^{+117.5}_{-11.6}$ & $-14.95 ^{+0.24}_{-0.19}$ \\
A2744\_z6\_33665 & 3.571744 & -30.411076 & 6.47$^{+0.39}_{-0.72}$ & 2.3$^{+0.6}_{-0.8}$ & $-17.31 ^{+0.55}_{-0.41}$ \\
A2744\_z6\_257 & 3.572845 & -30.380041 & 6.47$^{+0.35}_{-0.67}$ & 2.1$^{+0.7}_{-1.1}$ & $-16.80 ^{+0.52}_{-2.11}$ \\
M0416\_z6\_2011 & 64.051979 & -24.072380 & 6.47$^{+0.33}_{-0.32}$ & 1.8$^{+0.2}_{-0.6}$ & $-17.95 ^{+0.36}_{-0.31}$ \\
\enddata
\tablenotetext{a}{The errors given encompass the 68\% confidence interval of the photo-$z$.}
\tablenotetext{b}{The magnification given is the median and interquartile range from all lens models of the clusters updated after the Frontier Fields data were obtained.}
\tablenotetext{c}{The errors on the intrinsic magnitudes incorporate the errors on the measured fluxes and the magnification uncertainty.}
\end{deluxetable*}

\begin{deluxetable*}{l r r c c c}
\tablewidth{0pt}
\tablecaption{Catalog of the z$\sim$7 sample \label{tab:z7}}
\tablehead{\colhead{ID} & \colhead{RA} & \colhead{Dec} & \colhead{$z_{\mathrm{phot}}^a$} & \colhead{Magnification$^b$} & \colhead{$M_{1500}^c$} \\
 & \colhead{(J2000)} & \colhead{(J2000)} & & & \colhead{(intrinsic AB mag)}}
\startdata
M0416\_z7\_2844 & 64.017471 & -24.078640 & 6.29$^{+1.23}_{-0.42}$ & 2.2$^{+0.6}_{-0.7}$ & $-17.40 ^{+0.40}_{-0.32}$ \\
A2744\_z7\_3483 & 3.600199 & -30.406540 & 6.44$^{+0.88}_{-1.23}$ & 9.1$^{+2.7}_{-2.3}$ & $-15.22 ^{+0.45}_{-0.31}$ \\
A2744\_z7\_1922 & 3.582952 & -30.395275 & 6.49$^{+0.85}_{-1.01}$ & 14.0$^{+15.7}_{-6.6}$ & $-14.84 ^{+0.35}_{-0.60}$ \\
A2744\_z7\_3885 & 3.604148 & -30.409555 & 6.52$^{+0.41}_{-0.50}$ & 3.0$^{+1.8}_{-0.9}$ & $-16.77 ^{+0.34}_{-0.54}$ \\
A2744\_z7\_22919 & 3.604784 & -30.409237 & 6.56$^{+0.30}_{-0.26}$ & 2.7$^{+1.9}_{-1.1}$ & $-17.32 ^{+0.52}_{-0.44}$ \\
A2744\_z7\_955 & 3.576895 & -30.386328 & 6.58$^{+0.30}_{-0.37}$ & 4.2$^{+1.7}_{-1.7}$ & $-17.02 ^{+0.39}_{-0.71}$ \\
M0416\_z7\_2742 & 64.037743 & -24.077763 & 6.58$^{+0.51}_{-1.43}$ & 17.7$^{+141.8}_{-9.8}$ & $-14.49 ^{+0.30}_{-0.30}$ \\
M0416\_z7\_1860 & 64.028259 & -24.071136 & 6.59$^{+0.39}_{-0.48}$ & 4.9$^{+0.7}_{-1.3}$ & $-16.34 ^{+0.40}_{-0.36}$ \\
M0416\_z7\_1820 & 64.036591 & -24.070915 & 6.61$^{+0.47}_{-0.88}$ & 4.2$^{+4.4}_{-0.6}$ & $-16.34 ^{+0.49}_{-0.39}$ \\
A2744\_z7\_4485 & 3.593802 & -30.415449 & 6.63$^{+0.09}_{-0.09}$ & 3.5$^{+1.0}_{-1.4}$ & $-18.95 ^{+0.34}_{-0.39}$ \\
M0416\_z7\_4760 & 64.046059 & -24.094248 & 6.64$^{+0.26}_{-0.26}$ & 1.2$^{+0.1}_{-0.1}$ & $-18.61 ^{+0.20}_{-0.30}$ \\
M0416\_z7\_1176 & 64.032967 & -24.066185 & 6.64$^{+0.50}_{-0.48}$ & 5.9$^{+0.8}_{-2.3}$ & $-16.07 ^{+0.33}_{-0.42}$ \\
A2744\_z7\_203 & 3.580505 & -30.379473 & 6.66$^{+0.53}_{-1.14}$ & 2.9$^{+1.1}_{-0.9}$ & $-16.77 ^{+0.21}_{-0.20}$ \\
M0416\_z7\_3123 & 64.027702 & -24.080313 & 6.69$^{+1.34}_{-1.65}$ & 9.1$^{+5.5}_{-1.6}$ & $-15.24 ^{+0.42}_{-0.48}$ \\
A2744\_z7\_3258 & 3.580451 & -30.405041 & 6.70$^{+0.18}_{-0.12}$ & 4.5$^{+0.8}_{-1.0}$ & $-18.14 ^{+0.39}_{-0.34}$ \\
M0416\_z7\_2252 & 64.051277 & -24.074142 & 6.72$^{+0.65}_{-1.18}$ & 1.6$^{+0.2}_{-0.5}$ & $-17.28 ^{+0.33}_{-0.47}$ \\
A2744\_z7\_361 & 3.576530 & -30.381107 & 6.77$^{+0.38}_{-0.56}$ & 2.9$^{+0.6}_{-1.0}$ & $-16.94 ^{+0.41}_{-0.12}$ \\
A2744\_z7\_144455 & 3.598517 & -30.412600 & 6.79$^{+0.35}_{-0.69}$ & 12.5$^{+3.0}_{-6.7}$ & $-14.99 ^{+0.38}_{-0.51}$ \\
A2744\_z7\_524 & 3.577062 & -30.382597 & 6.79$^{+0.33}_{-0.57}$ & 2.5$^{+0.6}_{-1.0}$ & $-17.46 ^{+0.28}_{-0.22}$ \\
A2744\_z7\_3891 & 3.575346 & -30.409628 & 6.79$^{+0.62}_{-0.67}$ & 2.0$^{+0.3}_{-1.0}$ & $-16.92 ^{+0.35}_{-0.16}$ \\
A2744\_z7\_3345 & 3.600952 & -30.405600 & 6.81$^{+0.32}_{-0.53}$ & 4.6$^{+0.8}_{-1.1}$ & $-16.26 ^{+0.35}_{-0.15}$ \\
M0416\_z7\_4444 & 64.033630 & -24.090012 & 6.82$^{+0.50}_{-0.70}$ & 2.0$^{+0.5}_{-0.6}$ & $-17.99 ^{+0.48}_{-0.32}$ \\
M0416\_z7\_4848 & 64.043800 & -24.097269 & 6.86$^{+0.50}_{-0.76}$ & 1.2$^{+0.1}_{-0.2}$ & $-18.13 ^{+0.36}_{-0.55}$ \\
M0416\_z7\_4628 & 64.034615 & -24.092043 & 6.90$^{+0.62}_{-0.79}$ & 1.6$^{+0.3}_{-0.4}$ & $-17.24 ^{+0.25}_{-0.30}$ \\
M0416\_z7\_205 & 64.038391 & -24.057499 & 6.96$^{+0.65}_{-0.99}$ & 2.4$^{+0.3}_{-1.0}$ & $-17.00 ^{+0.34}_{-0.39}$ \\
M0416\_z7\_45828 & 64.034569 & -24.092089 & 6.96$^{+0.46}_{-0.22}$ & 1.6$^{+0.3}_{-0.4}$ & $-18.32 ^{+0.58}_{-0.35}$ \\
A2744\_z7\_3130 & 3.581287 & -30.404200 & 6.97$^{+0.59}_{-0.62}$ & 5.6$^{+6.2}_{-1.2}$ & $-15.96 ^{+0.38}_{-0.39}$ \\
A2744\_z7\_2250 & 3.585317 & -30.397957 & 7.02$^{+0.17}_{-0.13}$ & 3.6$^{+2.9}_{-1.2}$ & $-18.04 ^{+0.32}_{-0.47}$ \\
A2744\_z7\_974 & 3.579482 & -30.386545 & 7.04$^{+0.66}_{-0.40}$ & 4.8$^{+3.8}_{-2.1}$ & $-16.66 ^{+0.36}_{-0.16}$ \\
A2744\_z7\_2019 & 3.597845 & -30.395964 & 7.04$^{+0.16}_{-0.31}$ & 3.0$^{+2.2}_{-1.1}$ & $-18.27 ^{+0.37}_{-0.54}$ \\
A2744\_z7\_2700 & 3.579832 & -30.401588 & 7.12$^{+0.33}_{-0.89}$ & 8.2$^{+3.1}_{-1.7}$ & $-16.31 ^{+0.42}_{-0.27}$ \\
A2744\_z7\_497 & 3.598123 & -30.382389 & 7.33$^{+0.33}_{-1.19}$ & 1.5$^{+0.9}_{-0.5}$ & $-18.03 ^{+0.43}_{-0.52}$ \\
A2744\_z7\_22984 & 3.592275 & -30.409906 & 7.37$^{+0.23}_{-0.47}$ & 10.5$^{+9.3}_{-5.1}$ & $-17.23 ^{+0.38}_{-0.23}$ \\
A2744\_z7\_707 & 3.608731 & -30.384134 & 7.42$^{+0.35}_{-2.07}$ & 1.4$^{+0.7}_{-0.5}$ & $-17.98 ^{+0.48}_{-0.30}$ \\
A2744\_z7\_671 & 3.576698 & -30.383938 & 7.46$^{+0.42}_{-1.58}$ & 3.3$^{+0.5}_{-1.0}$ & $-17.07 ^{+0.28}_{-1.45}$ \\
A2744\_z7\_13225 & 3.570969 & -30.410618 & 7.48$^{+0.38}_{-1.55}$ & 2.2$^{+0.6}_{-0.7}$ & $-17.35 ^{+0.57}_{-0.29}$ \\
M0416\_z7\_114738 & 64.040985 & -24.084391 & 7.48$^{+0.35}_{-1.88}$ & 1.9$^{+0.4}_{-0.5}$ & $-17.26 ^{+0.65}_{-0.34}$ \\
A2744\_z7\_11040 & 3.581050 & -30.389601 & 7.52$^{+0.39}_{-1.53}$ & 14.1$^{+689.7}_{-5.3}$ & $-15.32 ^{+0.35}_{-0.60}$ \\
A2744\_z7\_545 & 3.608357 & -30.382893 & 7.59$^{+0.33}_{-2.02}$ & 1.7$^{+0.3}_{-0.8}$ & $-17.31 ^{+0.26}_{-0.13}$ \\
A2744\_z7\_82209 & 3.587596 & -30.399008 & 7.60$^{+0.31}_{-1.03}$ & 31.2$^{+23.1}_{-22.5}$ & $-15.43 ^{+0.37}_{-0.38}$ \\
\enddata
\end{deluxetable*}

\begin{deluxetable*}{l r r c c c}
\tablewidth{0pt}
\tablecaption{Catalog of the z$\sim$8 sample \label{tab:z8}}
\tablehead{\colhead{ID} & \colhead{RA} & \colhead{Dec} & \colhead{$z_{\mathrm{phot}}^a$} & \colhead{Magnification$^b$} & \colhead{$M_{1500}^c$} \\
 & \colhead{(J2000)} & \colhead{(J2000)} & & & \colhead{(intrinsic AB mag)}}
\startdata
A2744\_z8\_113946 & 3.585037 & -30.400955 & 7.47$^{+0.74}_{-0.71}$ & 46.2$^{+222.0}_{-41.4}$ & $-15.21 ^{+0.40}_{-0.32}$ \\
A2744\_z8\_398 & 3.605059 & -30.381470 & 7.54$^{+0.51}_{-1.19}$ & 1.3$^{+0.6}_{-0.5}$ & $-18.31 ^{+0.45}_{-0.31}$ \\
A2744\_z8\_1893 & 3.584220 & -30.395077 & 7.64$^{+0.52}_{-0.86}$ & 14.4$^{+50.4}_{-9.8}$ & $-15.68 ^{+0.35}_{-0.60}$ \\
M0416\_z8\_980 & 64.060333 & -24.064959 & 7.68$^{+0.32}_{-1.22}$ & 1.7$^{+0.0}_{-0.5}$ & $-18.15 ^{+0.34}_{-0.54}$ \\
A2744\_z8\_113489 & 3.587597 & -30.399065 & 7.78$^{+0.27}_{-0.45}$ & 67.5$^{+118.6}_{-57.3}$ & $-15.55 ^{+0.52}_{-0.44}$ \\
A2744\_z8\_482 & 3.603378 & -30.382254 & 8.16$^{+0.35}_{-0.24}$ & 1.4$^{+0.7}_{-0.5}$ & $-20.34 ^{+0.39}_{-0.71}$ \\
A2744\_z8\_110813 & 3.600142 & -30.383512 & 8.16$^{+0.31}_{-2.45}$ & 1.5$^{+0.8}_{-0.5}$ & $-17.49 ^{+0.30}_{-0.30}$ \\
A2744\_z8\_20572 & 3.573254 & -30.384312 & 8.16$^{+0.15}_{-1.99}$ & 2.6$^{+0.8}_{-1.1}$ & $-18.45 ^{+0.40}_{-0.36}$ \\
A2744\_z8\_417 & 3.606478 & -30.380989 & 8.19$^{+0.14}_{-0.50}$ & 1.8$^{+0.7}_{-0.4}$ & $-19.30 ^{+0.49}_{-0.39}$ \\
A2744\_z8\_892 & 3.596090 & -30.385834 & 8.23$^{+0.31}_{-0.16}$ & 1.9$^{+2.6}_{-0.5}$ & $-19.41 ^{+0.34}_{-0.39}$ \\
M0416\_z8\_12302 & 64.045082 & -24.078114 & 8.24$^{+0.30}_{-1.55}$ & 2.2$^{+0.4}_{-0.5}$ & $-18.03 ^{+0.20}_{-0.30}$ \\
M0416\_z8\_4209 & 64.037567 & -24.088114 & 8.36$^{+0.35}_{-0.65}$ & 1.9$^{+0.3}_{-0.5}$ & $-18.39 ^{+0.33}_{-0.42}$ \\
M0416\_z8\_2943 & 64.041412 & -24.078922 & 8.36$^{+1.02}_{-1.32}$ & 2.8$^{+0.3}_{-1.0}$ & $-17.16 ^{+0.21}_{-0.20}$ \\
M0416\_z8\_919 & 64.049591 & -24.064598 & 8.47$^{+0.14}_{-3.08}$ & 7.0$^{+11.3}_{-1.0}$ & $-16.97 ^{+0.42}_{-0.48}$ \\
M0416\_z8\_1341 & 64.040680 & -24.067390 & 8.47$^{+0.25}_{-1.86}$ & 5.8$^{+28.7}_{-2.0}$ & $-16.28 ^{+0.39}_{-0.34}$ \\
A2744\_z8\_130427 & 3.603858 & -30.382269 & 8.83$^{+****}_{-1.43}$ & 1.9$^{+0.7}_{-0.5}$ & $-19.35 ^{+0.33}_{-0.47}$ \\
\enddata
\end{deluxetable*}

\end{document}